\documentstyle[twocolumn,epsfig,aps]{revtex}
\begin{document}
\vspace{-2cm}
\title{The Interacting Gluon Model: a review}
\author{F.O. Dur\~aes$^{1,2}$\thanks{e-mail: fduraes@if.usp.br}, \
F.S. Navarra$^{2}$\thanks{e-mail: navarra@if.usp.br} \ and \ G.
Wilk$^{3}$\thanks{e-mail: Grzegorz.Wilk@fuw.edu.pl} \\[0.2cm]
{\it $^1$ Dep. de F\'{\i}sica, Faculdade de Ci\^encias Biol\'ogicas,
Exatas e Experimentais,} \\[0.1cm]
{\it Universidade Presbiteriana Mackenzie, C.P. 01302-907 S\~{a}o
Paulo,  Brazil}\\[0.1cm]
{\it $^2$Instituto de F\'{\i}sica, Universidade de S\~{a}o Paulo,}
\\[0.1cm]
{\it C.P. 66318, 05389-970 S\~{a}o Paulo, SP, Brazil} \\[0.1cm]
{\it$^3$ A. Soltan Institute for Nuclear Studies,
Nuclear Theory Department,}\\[0.1cm]
{\it 00-681 Warsaw, Poland}}
\maketitle
\vspace{1cm}

\begin{abstract}
The Interacting Gluon Model (IGM) is a tool designed to study energy
flow, especially stopping and leading particle spectra, in high energy
hadronic collisions. In this model, valence quarks fly through and
the gluon clouds of the hadrons interact strongly both in the soft
and in the  semihard regime. Developing this picture we arrive at a
simple description  of energy loss, given in terms of few parameters,
which accounts for a  wide variety of experimental data. This text is
a survey of our main  results and predictions.
\end{abstract}


\section{Introduction}

\subsection{Why a model?}

After more than 30 years of continuous advances we might expect
that Quantum Chromodynamics (QCD), the established theory of
strong interactions,  would provide us with a satisfactory
understanding of high energy hadronic  reactions. Unfortunately
this is not yet the case and we have to study these reactions
using models instead of the theory. This is so essentially because
of two reasons. The first one is because we can only perform
reliable calculations  in the perturbative regime, i.e., in
reactions where the momentum tranfer is larger than a few GeV.
However these represent only a small fraction of the  hadronic
cross sections. Indeed, even at very large energies most of events
involve low momentum transfer, as indicated by the average
transverse momentum of the produced particles, which is, in most
of the experiments, of the  order of 1 GeV or less. The second
reason is that the number of interacting  particles may be so
large that many body techniques and approximations are  needed. At
RHIC, for example, the number of finally produced hadrons may be
as large as 6000 (!) resulting from the complicated interaction
among a similar  number of quarks and gluons at the initial stage
of the collisions.  The study  of these systems can not be made
from first principles and models are required.

\subsection{What is a good  model?}

Since making models is inevitable and since there has been a
proliferation of models for particle production, we must try to
establish criteria to decide when a model is better than other. A
condition to be satisfied by a  model is a clear connection to the
underlying theory, i.e., the use of the  appropriate degrees of
freedom with the correct QCD interactions. Moreover,   assumptions
should be made only where the theory is not applicable and the
introduction of parameters should be restricted to a minimum.
Furthermore, a  model must have predictive power and be testable.
Even with these  constraints there are many implementations of the
basic QCD concepts and many  different ways to treat the 
non-perturbative dynamics.

Among all the existing models,  there are  some which try to give
a very comprehensive description of all possible experimental
data. Usually these models are at some stage transformed into
event generators and are used by experimental groups. While  they
may be helpful in projecting detectors and analyzing data, they
have the disadvantage of containing many parameters and of being 
a kind of "black box".  Well  known examples of
this type of model are HIJING \cite{wang}, VENUS \cite{wer1} and 
NEXUS \cite{wer2}.  A comprehensive list of 
available models of this kind can be found in \cite{GEN}. 
In a different approach there
are models which concentrate only on some more specific  features
both of theory and experiment. These models are more transparent,
easy to  handle, have only a small number of  parameters and are
devised to test only some limited   aspects of the theory. A famous
example of this kind of model is the thermal model, 
which is one of the first successful models formulated at a very
early stage of the research on multiparticle production \cite{Thermold}
and which remains very popular still in our days \cite{Thermnew}.
This model does
not involve amplitudes nor cross sections and is applied only to
systems which might be in thermal equilibrium. Typically we fit
simple thermal  distributions to the experimentally measured
transverse momentum spectra and extract the effective temperature.
Here we have little input and little output but we may learn
something studying different systems and, for example, establish  
 the behavior of the temperature as a function of the collision 
energy. The model
discussed here belongs to this cathegory of "economic" models. Our
aim is to decribe energy flow (stopping, energy deposition and
leading particle spectra) with a simple picture based on QCD, with
few parameters and  learn something from the analysis of data.

\subsection{Why study energy flow?}

Multiparticle production processes are the most complicated phenomena
as far as the number of finally involved degrees of freedom is
concerned. They also comprise the bulk of all inelastic collisions
and therefore are very important - if not {\it per se} then as a
possible background to some other, more specialized reactions
measured at high energy collisions. The large
number of degrees of freedom calls inevitably for some kind of
statistical or hydrodynamical descrition when addressing such
processes.  All corresponding models have to be supplemented with
information  about the fraction of the initial energy deposited in
the initial object  ("fireball") which is then the subject of further
investigations.  This fraction is called inelasticity and it is relevant 
also for low energy nuclear reactions \cite{subt}.

The knowledge of the energy deposited in the
central rapidity region in  heavy ion collisions at RHIC and LHC is
crucial \cite{QGP}.  Dividing this number by  the  volume of the formed 
system,  we will have an estimate of the initial  energy density  in such
collisions. If it is high enough we may be in a new phase of hadronic
matter: the plasma of quarks and gluons (QGP). 

On the other hand, the knowledge of the momentum spectrum of the
particles measured  in the large rapidity region, and, in particular,
those with the quantum numbers of the projectile (the so called
leading particles  or LP) gives valuable information about the 
non-perturbative dynamics of QCD.  Moreover, the LP spectrum 
and the inelasticity of the reaction are   
very useful in cosmic ray physics, in the description of the evolution 
of hadronic showers in the atmosphere \cite{Cosmic}. 

In the model considered here we hope to extract information about the
gluonic structure  of hadrons from observables like mass, diffractive
mass and leading particle spectra,  which are, at least in principle,
very easy to measure. This model describes only certain aspects of
hadronic collisions, related to energy flow and energy deposition in
the central rapidity region. It should not be regarded as an
alternative to a field-theoretical approach to amplitudes, but rather
as an extension of the naive parton model.  The reason for using it
is that it may be good  enough to account for energy flow in an
economic way. The deeper or more subtle aspects of the underlying
field theory  probably (this is our belief) do not manifest
themselves in energy flow, but rather in other quantities like  the
total cross section.  Inspite of its simplicity, this model can teach
us a few things and predict another few. This is encouraging because
in the near future new data from FERMILAB, RHIC and LHC  will be
available.

\subsection{A brief history of the IGM}

Long time ago, based on qualitative ideas advanced by Pokorski and
Van Hove \cite{Pokorski},  we started to develop a model to study
energy deposition,
connecting it with the apparent dominance of multiparticle production
processes by the gluonic content of the impinging hadrons, hence its
name: {\it Interacting Gluon Model} (IGM) \cite{IGM}. Its original
application to the  description of inelasticity \cite{Duraes93} and
multiparticle production processes in hydrodynamical treatments
\cite{Duraes94} was  followed by more refined applications to 
leading charm production \cite{Duraes95} and to  single
diffraction dissociation, both in hadronic reactions \cite{Duraes97a}
and in reactions originated by photons \cite{Duraes97b}. These works
allowed for providing the systematic description of the leading
particle spectra \cite{Duraes98a} and clearly demonstrated that
they are very sensitive to the amount of gluonic component in the
diffracted hadron as observed in  \cite{Duraes98b} and
\cite{Duraes98c}. We have found it remarkable that all the results
above were  obtained using the same set of basic parameters with
differences arising  essentially only because of the  different
kinematical limits present in each  particular application.  
All this points towards a kind of
{\it universality} of energy flow patterns in all the above mentioned
reactions. The IGM was further developed and fluctuations in impact 
parameter were included in \cite{HAMA}, where a careful study of the 
inelasticity in proton-nucleus reactions was performed. The model was 
employed by the Campinas group of cosmic ray physics to reanalyse
data from the AKENO collaboration and extract the proton-proton and 
proton-air cross sections \cite{godoi1}. This group used the IGM also 
to study the nucleonic and hadronic fluxes in the atmosphere 
\cite{godoi2}.

Recent experimental developments encouraged us to return  to the IGM
picture of energy flow. One of them was connected with the new, more
refined data on the leading proton spectra in $ep \rightarrow e'pX$
obtained recently by the ZEUS collaboration \cite{Garf03}.  Another one
was a recent work on  central mass production in Double Pomeron
Exchange (DPE) process  reported in \cite{Brandt02} allowing, in
principle,  for the extraction of the  Pomeron-Pomeron total cross
section $\sigma_{I\! P I\! P}$ (see \cite{Duraes02}).  Finally, in the 
last years it became possible to  study low $x$ physics experimentally, 
at HERA and
at RHIC. In this regime we probe the very low momentum region of the
gluon distributions in hadrons and nuclei, where a  qualitatively new
behavior is expected to be dominant. This newly explored sector of
the hadronic wave function is called  by some authors  Color Glass
Condensate  \cite{raju}. Its gluon density is so high that the
gluonic system can be treated  semi-classically and in the weak
coupling regime. This is the Bose-Einstein condensate  of the strong
interactions at the fundamental level and has been object of
experimental  searches. Some of its signatures are related to energy
flow observables and, in particular,   to the leading particle
spectrum. Indeed, in \cite{dumitru} it was suggested that the LP
spectrum in the saturation regime will go through a dramatically
softening. In some earlier  works we have predicted a slow softening
but we did not include saturation. In view of the  relevance of this
subject, we plan to address this problem in the near future. 

In the next section we shall provide a brief description of the IGM,
stressing  the universality of energy flow and then we devote the
other sections to discuss the  applications of the model.

\section{The model}

The IGM is based on the idea that
since about half of a hadron momentum is carried by gluons and since
gluons interact more strongly than quarks, during a high energy 
hadron-hadron collision there is a separation of constituents. 
Valence quarks tend to be fast forming leading particles whereas gluons 
tend to be stopped in the central rapidity region. The collision between 
the two gluonic clouds is treated as an
incoherent sum of multiple gluon-gluon collisions, the valence quarks
playing a secondary role in particle production. While this idea is
well accepted for large momentum transfer between the colliding
partons, being on the basis of some models of minijet and jet
production \cite{wang,trele,brown,gs,sj,geiger,lupia,sapeta}, in the IGM 
(and also  in \cite{brown} and \cite{sapeta})  its 
validity is extended down to low momentum transfers, only slightly larger 
than $\Lambda_{QCD}$. At first sight this is not  justified because at
lower scales there are no independent gluons, but rather a highly
correlated configuration of color fields. There are, however, some
indications coming from lattice QCD calculations, that these soft
gluon modes are not so strongly correlated. One of them is the result
obtained in \cite{gia}, namely that the typical correlation length of
the soft gluon fields is close to $0.3$ fm. Since this length is
still much smaller than the typical hadron size, the gluon fields
can, in a first approximation, be treated as uncorrelated. Another
independent result concerns the determination of the typical
instanton size in the QCD vacuum, which turns out to be of the order
of $0.3$ fm \cite{Shaefer98}. As it is well known (and has been
recently applied to high energy nucleon-nucleon and  nucleus-nucleus
collisions) instantons are very important as mediators of  soft gluon
interactions \cite{Shuryak00}. The  small size of the active
instantons lead to short distance  interactions between soft gluons,
which can be treated as independent.

These two results taken together give support to the idea that {\it a collision
between  two gluon clouds  may be viewed as a  sum of independent
binary gluon-gluon collisions}, which is the basic assumption of our model.
Developing the picture above with standard techniques  and enforcing
energy-momentum conservation, the IGM becomes  the ideal tool to
study energy flow in high energy hadronic collisions. Confronting  this
simple model with  several and different data sets we obtained a
surprisingly good agreement with experiment.

\begin{figure}[h]
\begin{center}
\centerline{\epsfig{figure=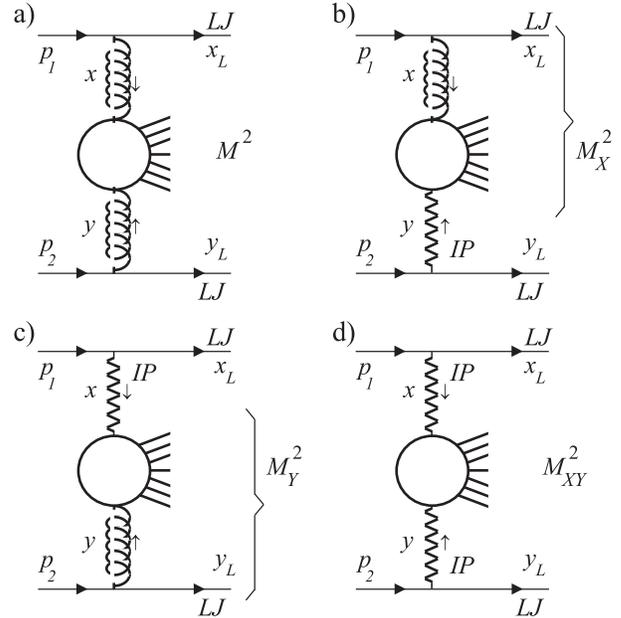,width=8.2cm}}
\caption{Schematic IGM pictures for $(a)$ non-diffractive (ND),
$(b)$ and $(c)$  single diffractive (SD) and $(d)$ double Pomeron
exchange (DPE) processes.} \label{igmscenarios}
\end{center}
\end{figure}

\vspace{-0.7cm}

\begin{figure}[h]
\begin{center}
\centerline{\epsfig{figure=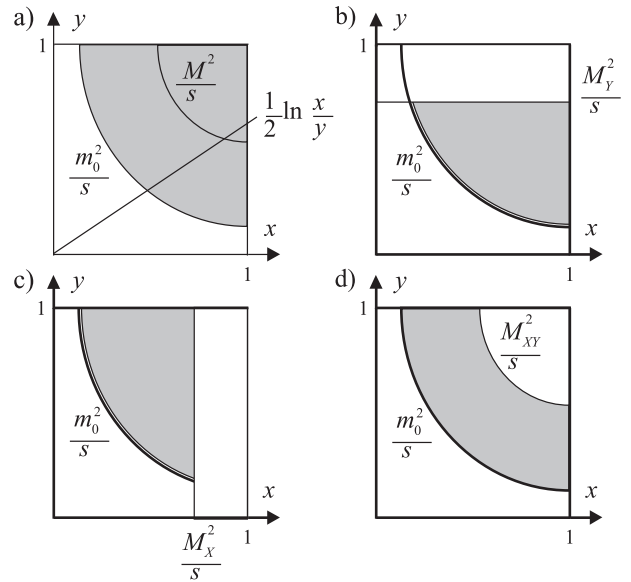,width=8.2cm}} \caption{Phase
space limits of ND, SD and DPE processes in the IGM. The
$\frac{1}{2}\ln \frac{x}{y}$ line in a) indicates the rapidity $Y$
of the produced mass $M$.} \label{igmphasespace}
\end{center}
\end{figure}

\vspace{-0.7cm}

The hadron-hadron  interaction follows the simple picture shown in
Fig. \ref{igmscenarios}: the valence quarks fly
through  essentially undisturbed whereas the gluonic clouds of both
projectiles  interact strongly with each other  forming a  central
fireball (CF) of mass $M$.  The two incoming projectiles $p_1$ and
$p_2$ loose fractions $x$ and $y$ of  their original momenta and get
excited forming what we call  leading jets (LJ's) carrying $x_L=
1-x$ and $y_L= 1-y$ fractions of the initial momenta. Depending on
the type of the process under consideration one  encounters the  different
situations depicted in Fig. \ref{igmscenarios}.

In a non-diffractive (ND) process (Fig. \ref{igmscenarios}a) one 
central fireball  of mass  $M$ is formed, whereas 
in single diffractive (SD) events
(Figs. \ref{igmscenarios}b and c) the corresponding diffractive systems 
have masses $M_X$ or $M_Y$ (comprising also the mass of CF). In 
double Pomeron exchanges (DPE) (Fig. \ref{igmscenarios}d) a   
CF of mass $M_{XY}$ is formed.  In Fig. \ref{igmphasespace} we show 
their corresponding phase space limits. The only difference between ND and 
SD or DPE processes is that in the latter cases the energy deposition is 
done by a restricted subset of gluons which in our language is  
 a  "kinematical" Pomeron ($I\!P$), the name which
we shall  use in what follows.

The central quantity in the IGM is $\chi(x,y)$, the probability to form a
CF carrying momentum fractions $x$ and $y$ of two colliding hadrons.
It follows from the quantitative implementation of the ideas
described above.  The essential ingredients are the assumption of
multiple independent gluon-gluon  collisions, low momentum dominance
of the gluon distributions and energy-momentum  conservation. The
derivation of our main formula, presented below, can be found in
Appendix A. $\chi(x,y)$ is given by:
\begin{eqnarray}
\chi (x,y) &=&\frac{\chi _0}{2\pi \sqrt{D_{xy}}}\,
\exp\{-\frac 1{2D_{xy}}
[ \langle y^2\rangle (x-\langle x\rangle )^2 \nonumber \\
&+&\, \langle x^2\rangle (y-\langle y\rangle )^2
+\,2\,\langle xy\rangle
(x-\langle x\rangle )(y-\langle y\rangle )] \}
\label{eq:CHI}
\end{eqnarray}
where $D_{xy} = \langle x^2\rangle \langle y^2\rangle -\langle
xy\rangle^2$ and
\begin{equation}
\langle x^n\, y^m\rangle =
\int_0^{x_{max}}\!dx'\,x'^n\,\int_0^{y_{max}}\!dy'\,y'^m\,\omega
(x',y'), \label{eq:defMOM}
\end{equation}
with $\chi _0$ defined by the normalization condition
\begin{equation}
\int_0^1\!dx\,\int_0^1\!dy\,\chi (x,y)\,\theta (x y - K_{min}^2) = 1\, ,
\end{equation}
where $K_{min}=\frac{m_0}{\sqrt{s}}$ is the minimal inelasticity
defined  by the mass $m_0$ of the lightest possible CF. 
The function $\omega(x',y')$, sometimes called the spectral
function, represents the average number of gluon-gluon collisions as
a function of  $x'$ e $y'$: 
\begin{eqnarray}
\omega(x',y')\,=\, \frac{d \overline {n}}{ d x' \, d y'} .
\label{omega1}
\end{eqnarray}
The appearance of the   number $\overline {n}$ comes from the 
use of  Poissonian distributions, which, in turn, is a consequence of
the assumption of independent  gluon-gluon collisions.    $\omega(x',y')$ 
contains all the dynamical inputs of the
model both  in the perturbative (semihard) and non-perturbative (soft)
regimes. The soft and semihard components are given by:
\begin{eqnarray}
\omega^{(S)} (x',y')&=&\frac{\hat{\sigma}^{(S)}
_{gg}(x'y's)}{\sigma (s)}\,G(x')\,G(y') \nonumber \\
&\times& \theta (x'y'-{K}_{min}^2) \,\, \theta 
(\frac{4\,{p_T}_{min}^2}{s} - x'y')
\label{eq:OMEGAS}
\end{eqnarray}
and
\begin{eqnarray}
\omega^{(H)} (x',y')&=&\frac{\hat{\sigma}^{(H)}
_{gg}(x'y's)}{\sigma (s)}\,G(x')\,G(y') \nonumber \\
&\times& \theta (x'y'-\frac{ 4\,{p_T}_{min}^2}{s})
\label{eq:OMEGAH}
\end{eqnarray}
where  $\hat{\sigma}^S _{gg}$ and $\hat{\sigma}^H _{gg}$ are the soft
and semihard gluonic cross sections, $p_{T_{min}}$ is the minimum
transverse momentum for minijet production and $\sigma$ denotes the
impinging projectiles cross section.

The values of $x_{max}$ and $y_{max}$ depend on the type of the
process  under consideration.    
For non-diffractive processes all phase space
contained in the shaded area  is allowed and in  this case we have:
\begin{equation}
x_{max}=y_{max}=1\,\,
\label{eq:cutnd}
\end{equation}

The effective number of gluons from the corresponding projectiles are
denoted by $G$'s and have been approximated in all our works by the
respective gluon distribution functions.  
There has been a remarkable progress in the knowledge of the parton 
distributions in hadrons \cite{cteq,mrst,dortmund}, especially in the 
low $x$ region, which becomes crucial at energies in the TeV range. 
Since in our previous applications 
of the IGM we have been studying collisions in the GeV domain, there was no 
need to use very  sophisticated parton distributions. Moreover, very often 
we needed parton densities at very low scales, which were not considered 
in the analyses presented in \cite{cteq,mrst,dortmund}. In some cases, we 
have used the parametrization of \cite{grv}, which is better suited for 
small scales. However,  as it will be shown, the IGM
can describe both the hadronic and nuclear collision data  with 
the following simple form of the gluon distribution function in the nucleon:
\begin{equation}
G(x) = p(m+1)\,\frac{(1-x)^m}{x}
\label{eq:gdx}
\end{equation}
with $m=5$ and the fraction of the energy-momentum allocated to
gluons is  equal to $p=0.5$.

In the IGM picture, diffractive and non-diffractive events have
been treated  on the same footing in terms of gluon-gluon
collisions. Single diffractive   processes receive great attention
mainly because of their potential ability to provide information
about the most important  object in the Regge theory, namely the
Pomeron ($I\!P$), its quark-gluon  structure and cross sections.
As can be seen in Fig. 1b (1c), the "diffractive  mass" $M_X$
($M_Y$) is just the invariant mass of a system composed of the CF
and LJ formed by one of the colliding projectiles. The main
difference with the  "non-diffractive mass" $M$ in Fig. 1a is that
the energy transfer from the  diffracted projectile is now done by
the highly correlated subset of gluons (denoted by $I\!P$) which
are supposed to be in a color singlet state. In  technical
terms it means that in comparison to the previous applications of
the  IGM cited before, we are free to change both the possible
shape of the function  $G(x)$ ($ \equiv G_{I\!P}(x) $),  the number of
gluons participating in the  process   
and the cross  section $\sigma$ ($\equiv \sigma_{p I\!P}$)  
in the spectral function $\omega$ used  above. 
The function  $ G_{I\!P}(x) $ should not be confused with the momentum  
distribution of the gluons inside the Pomeron, $f_{g/I\!P}(\beta)$. 
The  former  is given by the  convolution of the latter with the Pomeron 
flux factor as discussed in the Appendix B.    
Actually we have found that we can keep the shape of
$G(x)$ the same as before and the only change necessary  
to reach agreement with data is the amount of
energy-momentum $p=p_{I\!P}$ allocated  to the impinging hadron
and which will find its way in the object that we call $I\!P$.  It 
turns out that  $p_{I\!P} \simeq 0.05$, whereas 
$p\simeq 0.5$ for all gluons encountered so far. This choice,
with $m=5$ in eq. (\ref{eq:gdx}),  corresponds  to an intermediate
between ``soft'' and ``hard'' Pomeron (see Appendix B) and will
be used in what follows. Just in order to  make use of the present 
knowledge about the Pomeron, we have chosen  
$\sigma(s) = \sigma_{pI\!P} \, =
\, a + b \, \ln (s/s_0)$ where $s_0 = 1$ $GeV^2$  and $a = 2.6$
$mb$  and $b= 0.01$ $mb$.

In single diffractive processes only a limited part of
the phase  space supporting the $\chi(x,y)$ distribution is allowed
and in this case the integration  limits in the moments of the spectral 
function
$\omega$ (eq. (\ref{eq:defMOM}))  depend on the mass $M_X$ or $M_Y$
that is produced:
\begin{equation}
x_{max}=1\,\,\,;\,\,\,y_{max}=y\,\,\,;\,\,\,x_{max}\,
y_{max}=M_{X}^2/s  \label{eq:cutsd1}
\end{equation}
\begin{equation}
x_{max}=x\,\,\,;\,\,\,y_{max}=1\,\,\,;\,\,\,x_{max}\,
y_{max}=M_{Y}^2/s  \label{eq:cutsd2}
\end{equation}

By reducing these maximal values we select events in which the energy
released by the  projectile emitting $I\!P$ is small   
and at the same time  allow the formation of a rapidity gap
betwen the diffractive mass and the diffracted projectile. This  
is the experimental requirement defining a SD event.

Double Pomeron Exchange processes, inspite of their small cross
sections,  are inclusive measurements and do not involve particle
identification,  dealing only with energy flow. Such a process was
recently measured by UA8  \cite{Brandt02} and used to  deduce the 
$I\!PI\!P$ cross section,
$\sigma_{I\!PI\!P}$. It  turned out that using this method one gets
$\sigma_{I\!PI\!P}$ which  apparently depends on the produced mass
$M_{XY}$. This fact was tentatively  interpreted as signal of
glueball formation \cite{Brandt02}. In the IGM    a double
Pomeron exchange event (Fig. 1d) is seen as a specific type of energy
flow. The difference between it and the "normal" energy flow as
represented by Fig. 1a is that now the gluons involved in this
process must be confined to the object we called $I\!P$ above. We are
implicitly assuming that all gluons from $p_1$ and $p_2$
participating in the collision (i.e., those emitted from the upper
and lower vertex in Fig. 1d) have to form a color singlet. In this case 
two large rapidity gaps will form separating the diffracted hadron
$p_1$, the $M_{XY}$  system and the diffracted hadron $p_2$, which is
the experimental requirement  defining a DPE event. Also in this case
only a limited part of the phase  space supporting the $\chi(x,y)$
distribution is allowed and the limits  in the moments of the
spectral function $\omega$ (eq. (\ref{eq:defMOM}))  depend on the
mass $M_{XY}$ in the following way:
\begin{equation}
x_{max}=x\,\,\,;\,\,\,y_{max}=y\,\,\,;\,\,\,x_{max}\,y_{max}=M_{XY}^2/s
\, .
\label{eq:cutdpe}
\end{equation}

As before $G_{I\!P}(x)$ represent the number of gluons
participating in  the process and the cross section $\sigma$,
appearing in eqs. (\ref{eq:OMEGAS}) and (\ref{eq:OMEGAH}), represents
now the Pomeron-Pomeron cross section,  $\sigma_{I\!P I\!P}$.

The clear separation between valence quarks and bosonic degrees of 
freedom does not appear exclusively in the IGM. It appears also in soliton 
models of the nucleon \cite{diakonov}. In the Chiral Quark Soliton Model 
\cite{diakonov2}, for example, the nucleon is made of three massive quarks   
bound by the self-consistent pion field (the  "soliton").  It is interesting 
to observe that, according to this model, in a collision of two nucleons the 
valence quarks would interact much less than the pions and therefore would 
filter through and populate the large rapidity regions  leaving behind a blob 
of pionic matter in the central region.

\section{Inelasticity}

The energy dependence of inelasticity is an important problem
which is still subject of debate \cite{canal}. Generally speaking, 
inelasticity
$K$ is the fraction  of the total energy carried by the produced
particles in a given collision. However in the literature one
finds several possible ways to define it.  In the first one,
inelasticity is defined as
\begin{equation}
K_1 \, = \, \frac{M}{\sqrt{s}}
\label{eq.1nv}
\end{equation}
where $\sqrt{s}$ is the total reaction energy in its center of mass
frame and $M$ is the mass of the system (fireball, string, etc.)
which decays into the final produced particles. The second definition
of $K$ considered here is
\begin{equation}
K_2 \, = \, \frac{1}{\sqrt{s}} \, \sum_i \int dy \, \mu_i \,
         \frac{dn_i}{dy} \, \cosh y
\label{eq.2nv}
\end{equation}
where $\mu_i = \sqrt{p_{{\scriptscriptstyle T}_i}^2 + m^2_i}$ is the
transverse mass of the produced particles of type $i$ and $dn_i/dy$
their measured rapidity distribution. These two definitions are, in
principle, model independent, although the mass $M$ might be
difficult to evaluate in certain models.

The main difference between $K_1$ and $K_2$ is that, whereas the
first one refers to partons, the second one refers to final observed
hadrons. $K_2$ implicitly includes the kinetic energy of the object
of mass $M$. From the theoretical point of view, $K_1$  is a very
interesting quantity because it can be easy to calculate  and because
it is the relevant quantity when studying the formation  of dense
systems (e.g. quark-gluon plasma).

In Ref. \cite{Duraes93} we used the IGM to study
the energy dependence of $K_1$. We concluded that the introduction of
a semihard component (minijets) in that model produces increasing
inelasticities at the partonic level. In Ref. \cite{Duraes94} we
introduced a hadronization mechanism in the IGM, calculated the
rapidity distributions of the produced particles, compared our
results with the UA5 and UA7 data and finally calculated $K_2$. The
purpose of this exercise was to verify whether the hadronization
process changed our previous conclusion. {\it We found that, whereas
some  quantitative aspects, like the existence or not of Feynman
Scaling in  the fragmentation region and the numerical values of
$K_2$, depend very  strongly on details of the fragmentation process,
the statement that  minijets lead to increasing inelasticities
remains valid. }

\begin{figure}[h]
\begin{center}
\centerline{\epsfig{figure=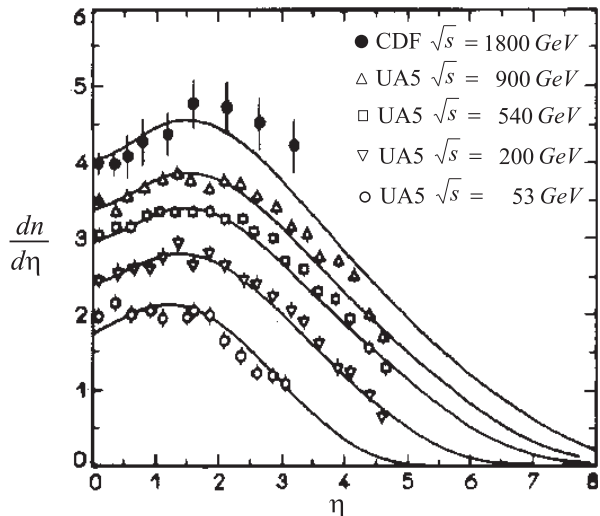,width=8cm}}
\caption{Pseudorapidity distributions measured at the central
rapidity  region. Data are from the UA5 collaboration
\protect\cite{GG} at different energies and from CDF collaboration
\protect\cite{Abe} at $\sqrt{s} = 1800 $  GeV. Full lines show the
IGM results.} \label{igmhadroni1}
\end{center}
\end{figure}

In Fig. \ref{igmhadroni1} we show the IGM pseudorapidity
distributions  compared to UA5 data at different energies
\cite{GG} and CDF \cite{Abe} data at $\sqrt{s} = 1800 $ GeV. For
the sake of comparison with other models  based both on soft and
semihard dynamics, we show in Fig. \ref{igmhadroni2}  our results
for the multiplicity (Fig. \ref{igmhadroni2}a) and central
rapidity  density (Fig. \ref{igmhadroni2}b) together with the
results of HIJING \cite{wang}  for the same quantities.


Both models fit the data but differ significantly when one switches
off the semihard (minijet) contribution. Whereas in HIJING  Feynman
Scaling violation in the  central region (the growth of \
${\displaystyle  \left. dn/d\eta \right|_{\eta=0}}$ \ with \
$\sqrt{s}\,$) is entirely  due to the minijets, in the IGM this
behavior is partly due to soft interactions, there being only a
quantitative difference when minijets are included.

In Fig. \ref{igminel} we plot $K_2$ (full lines) and $K_1$ (dashed
lines as a  function of $\sqrt{s}$. The lower curves show the results
when minijets are  switched off and only soft interactions take
place. The upper curves show the  effect of including minijets.

\begin{figure}[h]
\begin{center}
\centerline{\epsfig{figure=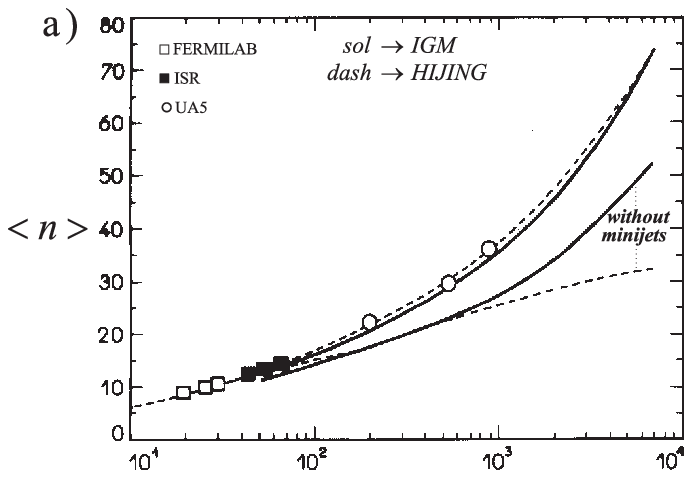,width=8cm}} 
\end{center}
\end{figure}

\vspace{-1.1cm}

\begin{figure}[h]
\begin{center}
\centerline{\epsfig{figure=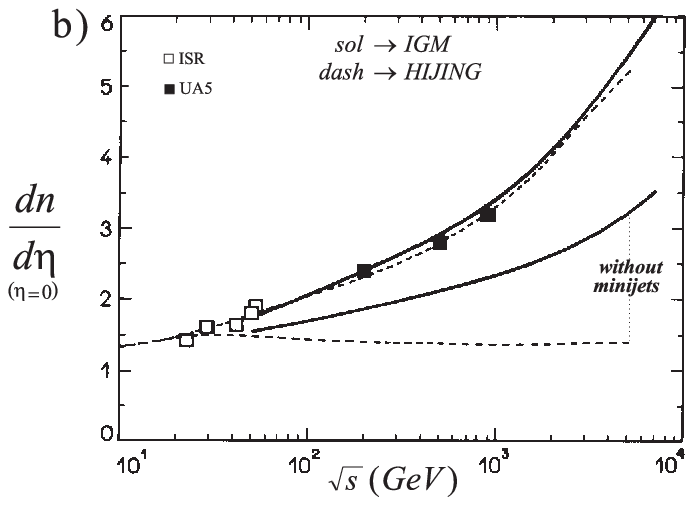,width=8cm}} \caption{a)
Average charged multiplicities as a function of the reaction
energy. Squares and circles are experimental data. Full lines show
the IGM results with and without the semihard contribution (lower
curve). Dashed lines show the same quantities calculated with
HIJING  \protect\cite{wang}. b) The same as a) for the central
pseudorapidity  distribution ${\displaystyle \left. dn/d\eta
\right|_{\eta=0}}$.} \label{igmhadroni2}
\end{center}
\end{figure}

\section{Leading charm and beauty}

In Ref. \cite{Duraes95} we treated leading charm production in
connection  with energy deposition in the central rapidity region
giving special attention  to the correlation between production in
central and fragmentation regions.  The significant difference
between the $x_F$ dependence of leading and nonleading charmed mesons  
\cite{Asymm}  
was not possible to be explained with the usual perturbative QCD
\cite{QCD} or with the string fragmentation model contained in PYTHIA
\cite{PYTHIA}.

\begin{figure}[h]
\begin{center}
\centerline{\epsfig{figure=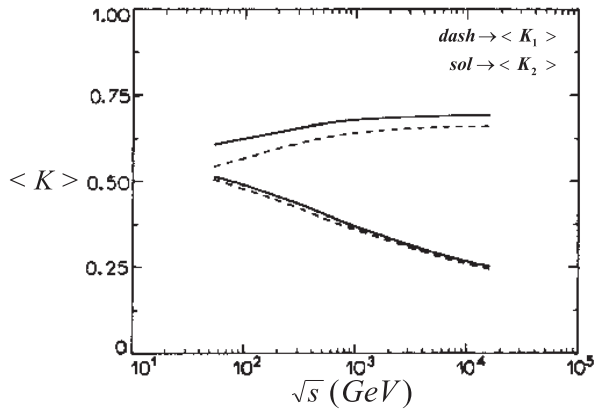,width=8.cm}}
\caption{Inelasticities $K_1$ (dashed lines) and $K_2$ (solid
lines) with minijets (upper curves) and without minijets (lower
curves) as a function of the reaction energy.} \label{igminel}
\end{center}
\end{figure}


In the case of pion-nucleon scattering, the measured leading charmed
mesons  \cite{Asymm} are $D^-$ and the nonleading are $D^+$. In the
spirit of IGM,  the central production ignores the valence quarks of
target and projectile   which, in the first approximation, just ``fly
through''.  Because of this, the centrally produced $D$'s do not show
any leading particle  effect. There are, however, two distinct ways
to produce $D$ mesons out of LJ's: fragmentation and recombination.
We assumed that, whenever energy allows,  we would have also
$\bar{c}c$ pairs in the LJ (produced, for example, from the  remnant
gluons present there). These charmed quarks might undergo
fragmentation  into $D$ mesons and also recombine with the valence
quarks. Whereas $D^+$ and  $D^-$ mesons are equally produced via
fragmentation only "leading" $D^-$'s  (which carry the valence quarks
of target and projectile) can be produced by  recombination. It turns
out that only this last process will produce asymmetry. 

The idea that $\bar{c}c$ pairs pre-exist in the projectiles can be made 
more precise and the origin of these ``intrinsic charm'' pairs can be 
attributed to the existence of a meson cloud around the nucleons and 
pions \cite{ma}.

The asymmetry in $D$ meson production can be defined as:
\begin{eqnarray}
A(x_F) &=& \frac{ \frac{d \sigma^{D^-}(x_F)}{d x_F} \,\,-\,\, 
\frac{ d \sigma^{D^+}(x_F)}{d x_F}} {\frac{d \sigma^{D^-}(x_F)}{d x_F} 
\,\, + \,\, \frac{d \sigma^{D^+}(x_F)}{d x_F}}
\label{assym}
\end{eqnarray}

\begin{figure}[h]
\begin{center}
\centerline{\epsfig{figure=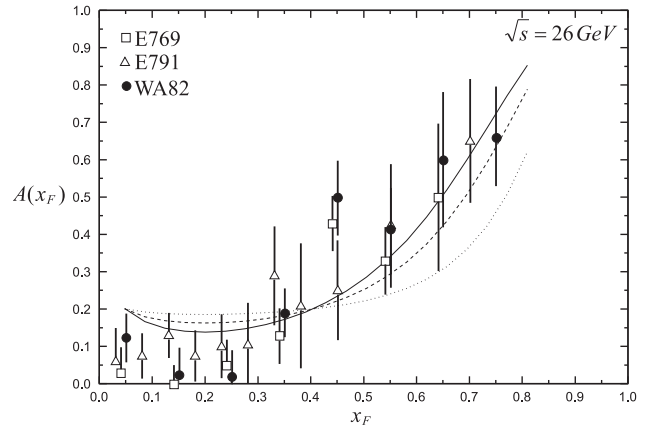,width=8.5cm}}
\caption{Asymmetry calculated with the IGM and compared with WA82
(solid circles),  with E769 (open squares) and E791 (open
triangles) data \protect\cite{Asymm}. Solid,  dashed and dotted
lines correspond to different  weights of the
recombination component equal to 80\%, 50\% and 20\% respectively.} 
\label{asymm1}
\end{center}
\end{figure}

\begin{figure}[h]
\begin{center}
\centerline{\epsfig{figure=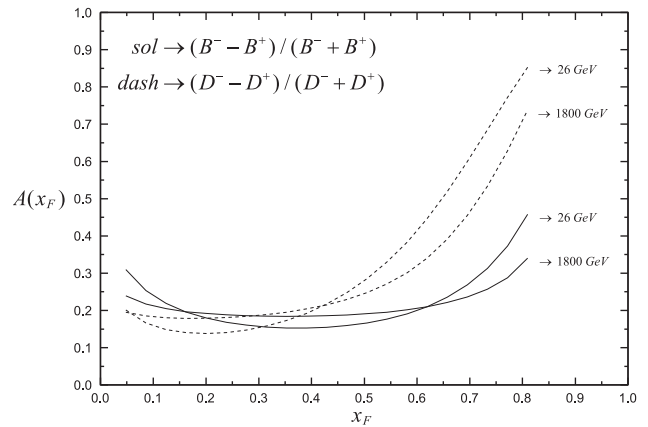,width=8.5cm}}
\caption{$B^-$/$B^+$ (solid lines) and $D^-$/$D^+$ (dashed lines)
asymmetries at $\sqrt{s}=$ 26  and 1800 GeV.} \label{asymm2}
\end{center}
\end{figure}

In Fig. \ref{asymm1} we compare our calculations with experimental
data from the  WA82, E769 and E791 \cite{Asymm} collaborations. {\it
The main conclusion of the work is that if one takes properly into
account  the correlation between energy deposition in the central
region and the leading particle momentum distribution, at higher
energies the increase of inelasticity  will lead to the decrease of
the asymmetry in heavy quark production.} In other words, if the
fraction of the reaction energy released in the central region
increases the asymmetry in the $x_F$ distributions of charmed mesons
will become smaller. In Fig. \ref{asymm2} we illustrate this
quantitatively and also consider  the leading beauty production.

\vspace{-0.7cm}

\section{Diffractive mass spectra in hadron-hadron collisions}

Diffractive scattering processes have received increasing
attention for several reasons. They are also related to the large
rapidity gap physics and the structure of the Pomeron. In a
diffractive scattering, one of the incoming hadrons emerges from
the collision only slightly deflected and there is a large 
rapidity gap between it and the other final state particles
resulted from the other excited  hadron.  Diffraction is due
to the Pomeron exchange but the exact nature of the Pomeron in QCD
is not yet  elucidated. The first test of a model of diffractive
dissociation  (SD) is the ability to properly describe the mass
($M_X$) distribution of diffractive systems, which has been
measured in many experiments \cite{Datadndmx} and parametrized  as
$(M_X^2)^{-\alpha}$ with $\alpha \simeq 1$.

In Ref. \cite{Duraes97a} we studied diffractive mass distributions
using the Interacting Gluon Model focusing on their energy dependence
and their connection with inelasticity distributions.  One advantage
of the IGM is that it was designed in such a way that the
energy-momentum conservation is taken care of before all other
dynamical  aspects. This feature makes it very appropriate for the
study of energy flow in high energy hadronic and nuclear reactions.
As mentioned before, in our  approach the definition of the object
$I\!P$ (see Fig. \ref{igmscenarios}b  and c) is essentially 
kinematical, very much in the spirit of those used  in other works
which deal with diffractive processes in the parton and/or  string
language. In order to regard our process as being  of the SD
type  we simply assume that all gluons from the target hadron
participating in the collision (i.e., those emitted from the lower
vertex in Fig. 1b)  have to  form a colour singlet. Only then a
large rapidity gap will form separating the diffracted  hadron and
the $M_X$ system.  Otherwise a colour string would develop, connecting the
diffracted hadron and the diffractive cluster, and would eventually
decay,  filling the rapidity gap with produced secondaries. As was
said above, it is a  special class of events in which the energy
released by the projectile emiting  $I\!P$ is small and consequently the
diffractive mass is small. Once only a limited part  of the phase
space is allowed, the limits in the moments of the spectral function
$\omega$ (eq. (\ref{eq:defMOM})), depend on the mass $M_X$ that is
produced through  the constraint $y_{max}=y=M_{X}^2/s$ (see eq.
(\ref{eq:cutsd1})).

\begin{figure}[h]
\begin{center}
\centerline{\epsfig{figure=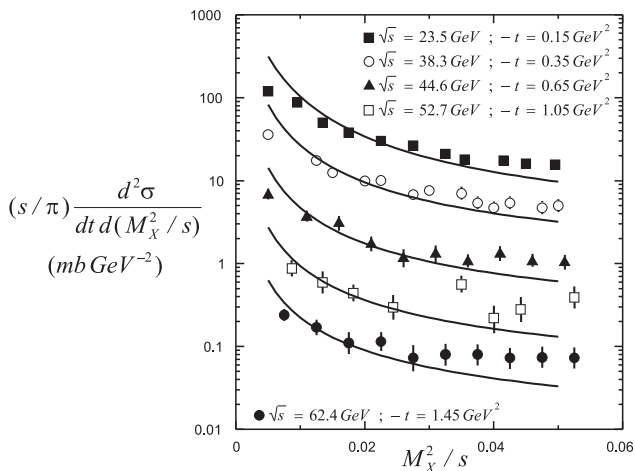,width=8.5cm}}
\caption{Diffractive mass spectrum for $pp$ collisions calculted
with the IGM  and compared with CERN-ISR data
\protect\cite{ISDATA}.} \label{dndmx1}
\end{center}
\end{figure}

\vspace{-0.7cm}

As shown in Appendix A, the mass spectra for SD processes is given
by:
\begin{eqnarray}
\frac{1}{\sigma}\frac{d\sigma}{dM_X^2}&=&\frac{dN}{dM_X^2}=
\int_0^1dx\,\int_0^1dy\,\chi(x,y)\nonumber \\
&\times&\,\delta\left(M^2_X-sy\right)\,\theta\left(xy -
K_{min}^2\right) \label{eq:SD}
\end{eqnarray}


Although in the final numerical calculations the above complete
formulation is used, it is worthwhile to present approximate
analytical results in order to illustrate the main characteristic
features of the IGM diffractive dissociation processes. By keeping
only the most singular terms in gluon distribution functions,
i.e., $G(x)\simeq \frac{1}{x}$ and only the leading terms in
$\sqrt{s}$, as shown in Appendix A, we arrive at the following
expression:
\begin{eqnarray}
\frac{dN}{dM_X^2}\, &\simeq&\, \frac{1}{s}\,
H(M_X^2,s)\,
                   F(M^2_X,s) \nonumber\\
              &\simeq&\, \frac{\rm const}{M_X^2}\,
                        \frac{1}{\sqrt{c\, \ln\frac{M_X^2}{m_0^2}}} \,
                        \exp\left[ - \frac{\left(1\, -\,
                        c\, \ln\frac{M_X^2}{m_0^2}\right)^2}
                        {c\, \ln\frac{M_X^2}{m_0^2}}\right].
                        \label{eq:SDaprox}
\end{eqnarray}
where $c$ is a constant and $m_0$ is a soft energy scale. 
The expression above is governed by the $\frac{1}{M^2_X}$ term.
The other two terms have a weaker dependence on $M^2_X$. They
distort the main ($\frac{1}{M^2_X}$) curve in opposite directions
and tend to compensate each other. It is therefore very
interesting to note that even before choosing a very detailed form
for the gluon distributions and hadronic cross sections we obtain
analytically the typical shape of a diffractive spectrum,
$\frac{1}{M^2_X}$.

\begin{figure}[h]
\begin{center}
\centerline{\epsfig{figure=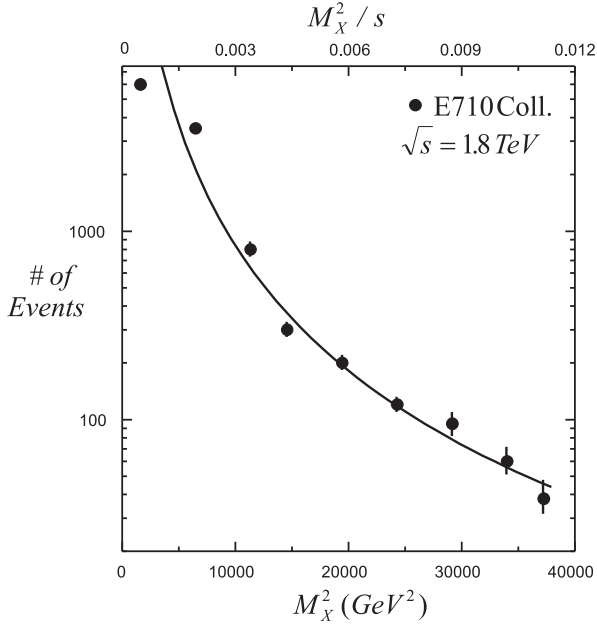,width=8cm}}
\caption{Diffractive mass spectrum for $p\bar{p}$ collisions
calculted with the IGM and compared with FERMILAB Tevatron data
\protect\cite{E710}.} \label{dndmx2}
\end{center}
\end{figure}

In Fig. \ref{dndmx1} we show our diffractive mass spectrum and
compare it to experimental data from the CERN-ISR \cite{ISDATA}. Fig.
\ref{dndmx2} shows the  diffractive mass spectrum for $\sqrt{s}=
1800$ GeV compared to experimental data  from the E710 Collaboration
\cite{E710}. In these curves we have used our  intermediate Pomeron
profile: $G_{I\!P}(y)$ given by (\ref{eq:gdx}) with $m=5$  and
$p_{{I\!P}}=0.05$.

\begin{figure}[h]
\begin{center}
\centerline{\epsfig{figure=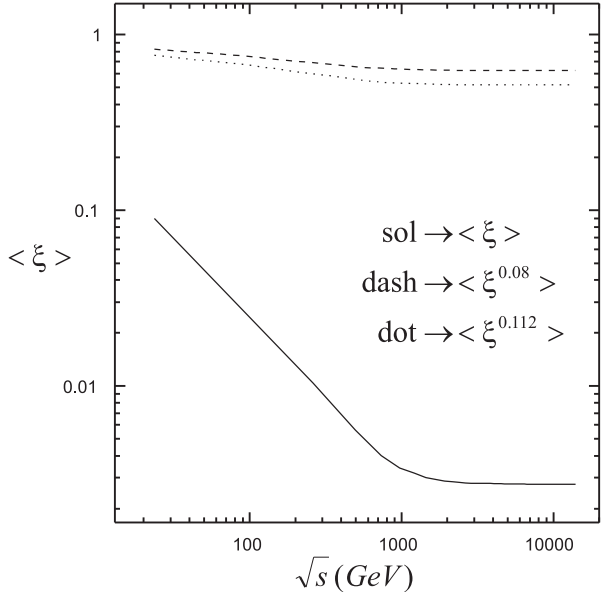,width=8cm}}
\caption{Energy dependence of the "diffractive inelasticity"
$\langle \xi \rangle$ and of $\langle \xi^{\varepsilon} \rangle$.}
\label{xids}
\end{center}
\end{figure}

In all curves we observe a modest narrowing as the energy increases. This small
effect means that the diffractive mass becomes a smaller fraction of the available
energy $\sqrt{s}$. {\it In other words, the "diffractive inelasticity" decreases with
energy and consequently the "diffracted leading particles" follow a harder $x_F$
spectrum}. Physically, in the context of the IGM, this means that the deposited
energy is increasing with $\sqrt{s}$ but it will be mostly released outside the
phase space region that we are selecting. A measure of the "diffractive inelasticity"
is the quantity  $\xi=M^2_X/s$. It is very simple to calculate its average
value $\langle \xi \rangle$ from the diffractive mass spectrum. Making a trivial
change of variables we get:
\begin{eqnarray}
\langle \xi \rangle \left( s \right) &=& \int_{\xi_{min}}^{\xi_{max}}
d\xi\, \frac{d N}{d \xi}\,  \xi
\label{eq:XI}
\end{eqnarray}
where $\xi_{min}$ ($=1.5/s$) and $\xi_{max}$ ($= 0.1$) are the
same used in other works. In Fig. \ref{xids} we plot $\langle \xi
\rangle $  against $\sqrt{s}$. As it can be seen $\langle \xi
\rangle$ decreases with  $\sqrt{s}$ not only because $\xi_{min}$
becomes smaller but also because  $d N/d \xi$ changes with the
energy, falling faster. Also shown in Fig. \ref{xids} is the quantity
$\langle \xi^{\varepsilon} \rangle$ (sometimes used  in connection
with the energy dependence of the single diffractive cross-section)
for $\varepsilon= 0.08$ (dashed lines) and $\varepsilon= 0.112$
(dotted lines).

\vspace{-0.7cm}

\vspace{-0.7cm}

\section{Diffractive mass spectra in electron-hadron collisions}

At the HERA electron-proton collider the bulk of the cross section
corresponds  to photoproduction, in which a beam electron is
scattered through a very small  angle and a quasi-real photon
interacts with the proton. For such small virtualities  the dominant
interaction mechanism takes place via fluctuation of the photon into
a  hadronic state (vector meson dominance) which interacts with the
proton via the  strong force. High energy photoproduction therefore
exhibits similar characteristics  to hadron-hadron interactions.

In Ref. \cite{Duraes97b} we studied diffractive mass distributions in
a photon-proton  collision. The photon is converted into a mesonic
state and then interacts with the  incoming proton. The diffractive
meson-proton interaction follows then the usual IGM  picture. The
diffracted proton  in Fig. \ref{igmscenarios}b), looses only a
fraction $y$ of its momentum but otherwise remains intact. In the
limit $y\rightarrow 1$, the whole available energy is stored in $M_X$
which then remains at rest, i.e., $Y_X = 0$. For small values of $y$
we have small masses $M_X$ located at large rapidities $Y_X$. As
before the upper cut-off $y_{max}$ ($=y=M_X^2/s$) is a  kinematical
restriction preventing the gluons coming from the diffracted proton
(and forming our object $I\!P$) to carry more energy than what is
released in the diffractive system. It plays a central role in the
adaptation of the IGM to  diffractive dissociation processes being
responsible for its proper $M^2_X$  dependence.


\begin{figure}[h]
\begin{center}
\centerline{\epsfig{figure=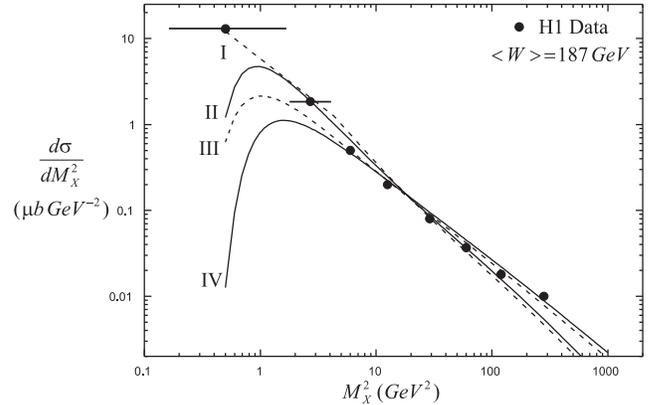,width=8.5cm}}
\caption{Diffractive mass spectrum for $\gamma p$ collisions at
$W=187\,GeV$ calculted with the IGM and compared with H1 data
\protect\cite{HERA}. The different curves  correspond to the
choices: I ($m_{0}=0.31\,GeV$, $\sigma=2.7\,mb$),  II
($m_{0}=0.35\,GeV$, $\sigma=2.7\,mb$),  III ($m_{0}=0.31\,GeV$,
$\sigma=5.4\,mb$) and IV ($m_{0}=0.35\,GeV$, $\sigma=5.4\,mb$),
respectively.} \label{dndmxgamap1}
\end{center}
\end{figure}

In the upper leg of Fig. \ref{igmscenarios}b) we have assumed, for
simplicity,  the vector meson to be $\rho^{0}$ and take $
G^{\rho^{0}}(x) = G^{\pi}(x)$.  Since the parameter $p/\sigma$
appearing in eqs. (\ref{eq:OMEGAS}) and (\ref{eq:OMEGAH}) has been
fixed considering the proton-proton diffractive dissociation and here
we are  addressing the $p\rho^{0}$ case there exists some freedom to
change $\sigma$.  We can also investigate the effect of small changes
in the value of $m_{0}$ on  our final results.

\begin{figure}[h]
\begin{center}
\centerline{\epsfig{figure=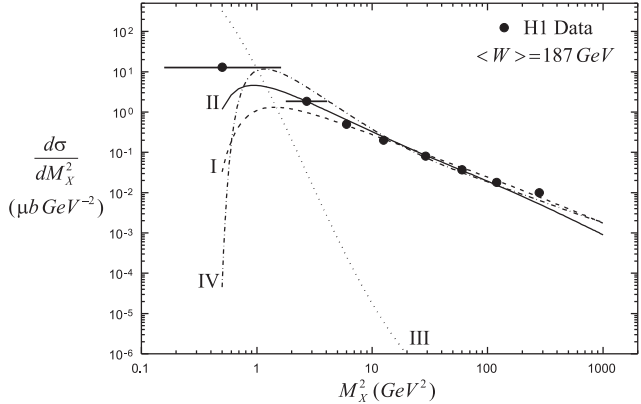,width=8.5cm}}
\caption{Diffractive mass spectrum for $\gamma p$ collisions at
$W=187\,GeV$ calculted with the IGM and compared with H1 data
\protect\cite{HERA}. The solid line (curve II) corresponds to the
choice $m_{0}=0.35\,GeV$, $\sigma=2.7\,mb$ and $G_{I\!P}(y)$.
Curves I (dashed) and III (dotted) are obtained replacing
$G_{I\!P}(y)$ by $G^h_{I\!P}(y)$ and $G^s_{I\!P}(y)$ respectively.
Curve IV is obtained with $G^s_{I\!P}(y)$ and $m_{0}=0.50\,GeV$
and $\sigma=5.4\,mb$.} \label{dndmxgamap2}
\end{center}
\end{figure}

\vspace{-0.7cm}

In Fig. \ref{dndmxgamap1} we compare our results, eq. (\ref{eq:SD}),
for different  choices of $m_{0}$ and $\sigma$ with the data from the
H1 collaboration \cite{HERA}.  In all these curves we have used our
intermediate Pomeron profile.  In Fig. \ref{dndmxgamap2}  we
compare the same data with our mass spectrum obtained with
$G^h_{I\!P}(y)$  (curve I), $G_{I\!P}(y)$ (curve II) and
$G^s_{I\!P}(y)$ (curve III). {\it This comparison suggests that the
``hard'' Pomeron can give a good description of  data}. The same can
be said about our ``mixed'' Pomeron, which, in fact seems  to be more
hard than soft. These three curves were calculated with exactly the
same parameters and normalizations, the only difference being the
Pomeron profile. Soft and hard gluon distributions
$G^{s,h}_{I\!P}(y)$ are calculated  in the Appendix B.  
{\it Apparently the ``soft'' Pomeron (curve III) is ruled out by
data. }

\section{Diffractive mass spectra in double Pomeron exchange}

After ten years of work at HERA, an impressive amount of knowledge
about the  Pomeron has been accumulated, especially about its
partonic composition and  parton distribution functions. Less known
are its interaction properties.  Whereas the Pomeron-nucleon cross
section has been often discussed in the  literature, the recently
published data by the UA8 Collaboration \cite{Brandt02}  have shed
some light on the Pomeron-Pomeron interaction. In \cite{Brandt02} the
Double Pomeron Exchange cross section was written as the product of
two  flux factors with the $I\!PI\!P$ cross section, $\sigma_{I\!P
I\!P}$,  being thus directly proportional to this quantity. This
simple formula relies  on the validity of the Triple-Regge model, on
the universality of the Pomeron  flux factor and on the existence of
a factorization formula for DPE processes.  However, for these
processes the factorization hypothesis has not been proven  and is
still matter of debate \cite{collins,bersop,bercol,terron}. In
\cite{berera}  it was shown that factorizing and non-factorizing DPE
models may be experimentally  distinguished in the case of dijet
production.

Fitting the measured  mass spectra allowed for the determination of
$\sigma_{I\!P I\!P}$ and its dependence on $M_X$, the mass of the
diffractive  system. The first observation of the UA8 analysis was
that the measured  diffractive mass ($M_X$) spectra show an excess at
low values  that can hardly  be explained with a constant (i.e.,
independent of $M_X$) $\sigma_{I\!P I\!P}$.  Even after introducing
some mass dependence in $\sigma_{I\!P I\!P}$ they were  not able to
fit the spectra in a satisfactory way. Their conclusion was that  the
low $M_X$ excess may have some physical origin like, for example
glueball  formation.

Although the analysis performed in \cite{Brandt02} is standard, it is
nevertheless  useful to confront it with the IGM 
description of the diffractive  interaction. Double Pomeron exchange
processes, inspite of their small cross  sections, appear to be an
excellent testing ground for the IGM because they are  inclusive 
measurements and do not involve particle identification, dealing only
with energy flow. In Ref. \cite{Duraes02} we studied the diffractive
mass distribution observed  by UA8 Collaboration in the inclusive
reaction $p\bar p \rightarrow p X \bar p$ at  $\sqrt s = 630 \,GeV$,
using the IGM with DPE included. The interaction follows the
picture shown in Fig. \ref{igmscenarios}d.

As shown  in Appendix A, the mass spectrum for DPE processes is given
by:
\begin{eqnarray}
\frac{1}{\sigma}\frac{d\sigma}{dM_{XY}}&=&\frac{dN}{dM_{XY}}=
\int_0^1 dx\,\int_0^1 dy\,\chi(x,y) \nonumber \\
&\times&\,\delta\left(M_{XY}-\sqrt{xys}\right)\,\theta\left(xy -
K_{min}^2\right) \label{eq:DPE}
\end{eqnarray}

As indicated in the recent literature
\cite{collins,bersop,bercol,terron,berera}, one of the crucial issues
in  diffractive physics is the possible breakdown of factorization.
As stated in \cite{bersop} one may have Regge  and hard
factorization. Our model does not rely on any of them.  In the
language used in \cite{bersop}, we  need and use a ``diffractive
parton distribution'' and we do not really need to talk about ``flux
factor'' or ``distribution of partons  in the Pomeron''. Therefore
there is no Regge  factorization implied. However, we will do this
connection in eq. (\ref{conv}) of the Appendix B, in order to make
contact with the Pomeron pdf's parametrized by the H1 and ZEUS
collaborations. As for hard factorization, it is valid as long as the
scale $\mu$ is large. In the IGM, as it will be seen,  the scale is
given by  $\mu^2 = x y s$, a number which  sometimes is larger than
$3 - 4 \, GeV^2$ but sometimes is smaller, going down to values only
slightly above $\Lambda_{QCD}^2$. When the scale is large ($ \mu^2 >
p_{T_{mim}}^2$)  we employ Eq. (\ref{eq:OMEGAH})  and  when it is
smaller ($m^2_0 < \mu^2 < p_{T_{mim}}^2$) we use  Eq.
(\ref{eq:OMEGAS}). Therefore, in part of the phase space we are
inside the validity domain of hard   factorization, but very often we
are outside this domain. From the practical point of view,  Eq.
(\ref{eq:OMEGAH}), being defined at a semihard scale, relies on hard
factorization for  the elementary $gg \rightarrow gg$ interaction,
uses parton distribution function extracted from DIS  and an
elementary cross section $\hat{\sigma}_{gg}$ taken from standard pQCD
calculations.  The validity of the factorizing-like formula Eq.
(\ref{eq:OMEGAS}) is  {\it an assumption of the model}. In fact, the
relevant scale there is $m_0^2 \simeq \Lambda_{QCD}^2$ and, strictly
speaking,  there are no rigorously defined parton distributions,
neither elementary cross sections.  However, using  Eq.
(\ref{eq:OMEGAS}) has non-trivial consequences which were in the past
years supported by an extensive  comparison with experimental data.
In this approach, since we have fixed all parameters  using previous
data on leading particle formation and single diffractive mass
spectra,  there are no free parameters here, except $\sigma_{I\!P
I\!P}$.

\begin{figure}[h]
\begin{center}
\centerline{\epsfig{figure=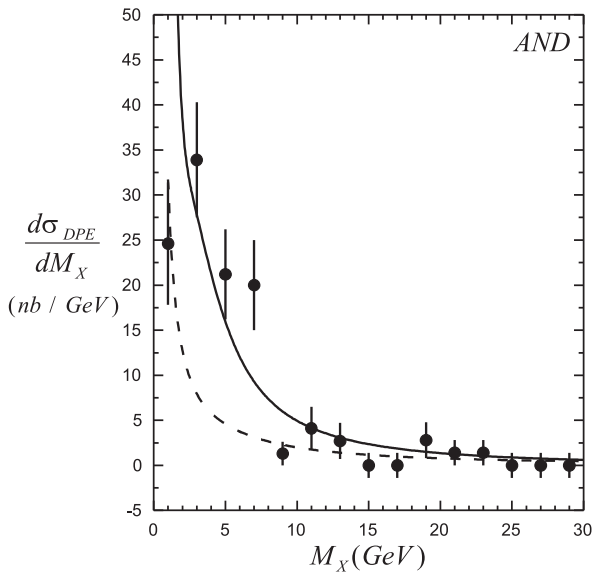,width=8cm}} \caption{IGM
DPE diffractive mass distributions: Solid and dashed lines show
the  results with $\sigma_{I\!P I\!P}$ equal to $0.5\,mb$ and
$1.0\,mb$, respectively,  calculated with the intermediate Pomeron
profile. Our curves were normalized to the ``AND" data sample of
\protect\cite{Brandt02}.} \label{dndmxdpe1}
\end{center}
\end{figure}

We start evaluating Eq. (\ref{eq:DPE}) with the inputs that were
already fixed  by other applications of the IGM, namely,
(\ref{gFeynman}) with $p_d=0.05$. In  Fig. \ref{dndmxdpe1} we show
the numerical results for DPE mass distribution normalized  to the
``AND" data sample of \cite{Brandt02}. We have fixed  the
parameter $\sigma$ ($\equiv \sigma_{I\!\!P I\!\!P}$) appearing in
Eq. (\ref{eq:OMEGAS}) and Eq. (\ref{eq:OMEGAH}), to $0.5\,mb$
(dashed lines) and $1.0\,mb$ (solid lines).

We emphasize that, in this approach, since we have fixed all
parameters using previous  data on leading particle formation and
single diffractive mass spectra, there are no free parameters
here, except $\sigma_{I\!\!P I\!\!P}$. {\it As it can be  seen
from the figure, in our model we obtain the fast increase of
spectra in the low mass region without the use of a $M_X$
dependent $I\!\!P$ $I\!\!P$ cross section and  this quantity seems
to be approximately $\sigma_{I\!\!P I\!\!P} \simeq 0.5\,mb$. }

\vspace{-0.7cm}

\begin{figure}[h]
\begin{center}
\centerline{\epsfig{figure=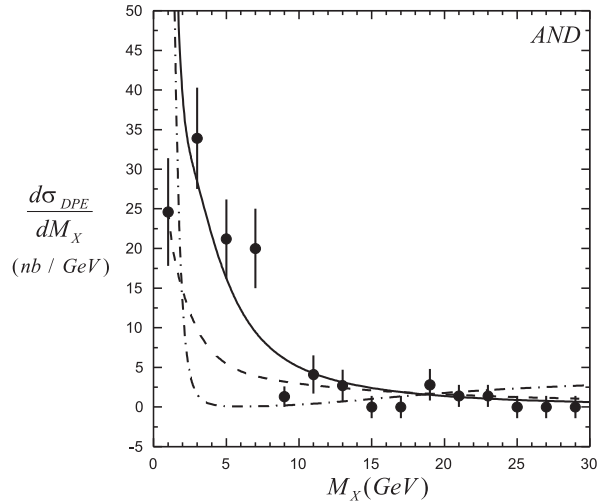,width=8cm}} \caption{IGM
DPE diffractive mass distributions: Solid line as in Fig.
\ref{dndmxdpe1}, dashed and dash-dotted lines represent the
``hard" and  ``super-hard" Pomeron profiles. In all cases
$\sigma_{I\!P I\!P}=0.5\,mb$.  Our curves were normalized to the
``AND" data sample of \protect\cite{Brandt02}.} \label{dndmxdpe2}
\end{center}
\end{figure}

\vspace{-0.7cm}

We next replace (\ref{gFeynman}) by the convolution (\ref{conv})  to
see  which of the previously considered Pomeron profiles, hard or
superhard, gives the best fit of the  UA8 data. In doing so, we shall
keep everything else the same, i.e., $p_d= 0.05$ and  $\sigma_{I\!\!P
I\!\!P} = 0.5 \, mb$.

In Fig. \ref{dndmxdpe2}, we repeat the fitting procedure used in
Fig. \ref{dndmxdpe1}  for these Pomeron profiles. Solid, dashed
and dash-dotted lines represent respectively  Eq.
(\ref{gFeynman}), hard and superhard Pomerons. We see that, for
harder Pomeron  profiles we ``dig a hole'' in the low mass region
of the spectrum. Note that the solid  lines are the same as in
Fig. \ref{dndmxdpe1}. Looking at the figure, at first sight, we
might be tempted to say that Eq. (\ref{gFeynman}) gives the best
agreement with data and  a somewhat worse description can be
obtained with the hard Pomeron (in dashed lines),  the superhard
being discarded. However, comparing the dashed lines in  Fig.
\ref{dndmxdpe1} and Fig. \ref{dndmxdpe2} and observing that they
practically  coincide with each other, we conclude that the same
curve can be obtained either with   (\ref{gFeynman}) and
$\sigma_{I\!\!P I\!\!P} = 1.0 \, mb$ (dashed line in  Fig.
\ref{dndmxdpe1}) or with (\ref{conv}), (\ref{fhard}) and
$\sigma_{I\!\!P I\!\!P} =  0.5 \, mb$ (dashed line in Fig.
\ref{dndmxdpe2}).  {\it In other words we can trade the
``hardness'' of the Pomeron with its interaction  cross section.
The following two objects give an equally good description of
data: i) a Pomeron composed by more and softer gluons and with a
larger cross section and ii) a Pomeron made by fewer, harder
gluons with a smaller interaction cross section.} We have checked
that this reasoning can be extended to the superhard Pomeron.
Although, apparently disfavoured by Fig.  \ref{dndmxdpe2}
(dash-dotted lines), it might still fit the data provided that
$\sigma_{I\!\!P I\!\!P} < 0.25 \, mb$. Given the uncertainties in
the data  and the limitations of the model, we will not try for
the moment to refine this  analysis. It seems possible to describe
data  in  a number of different ways. {\it We conclude then that
nothing exotic has been observed and also that the Pomeron-Pomeron
cross section is bounded to be $\sigma_{I\!\!P I\!\!P} < 1.0 \,
mb$. }

\begin{figure}[h]
\begin{center}
\centerline{\epsfig{figure=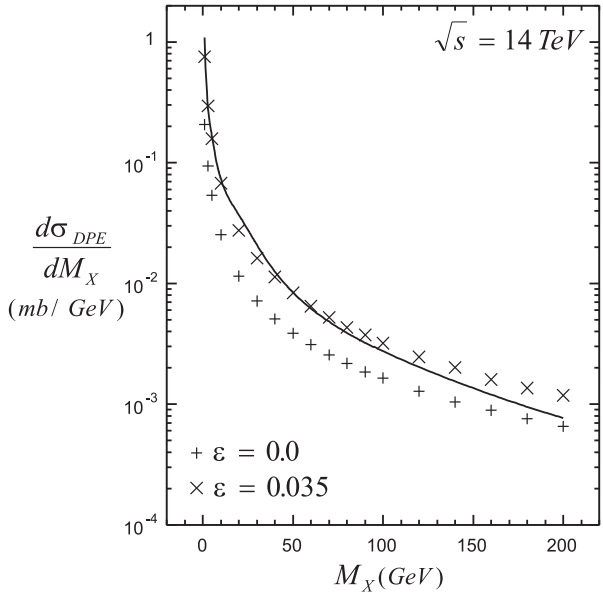,width=8cm}}
\caption{IGM
prediction for $d\sigma/dM_X$ at LHC with $\sigma_{I\!P
I\!P}=1.0\,mb$.  Cross($+$) and Cross($\times$) are predictions
made by Brandt {\it et al.} \protect\cite{Brandt02} for two values
of effective Pomeron intercepts $\alpha (0) = 1+\varepsilon$.}
\label{dndmxdpe3}
\end{center}
\end{figure}

\vspace{-0.7cm}

In Fig. \ref{dndmxdpe3} we compare our predictions for
$d\sigma/dM_X$ (mb/GeV)  for LHC ($\sqrt s = 14$ TeV) assuming an
$M_X$-independent $\sigma_{I\!\!P I\!\!P}=1.0\,mb$ (and using
(\ref{gFeynman})) with predictions made  by Brandt {\it et al.}
\cite{Brandt02} for two values of effective Pomeron intercepts
($\alpha (0) = 1+\varepsilon$), $\varepsilon = 0.0 $ and $0.035$.

\begin{figure}[h]
\begin{center}
\centerline{\epsfig{figure=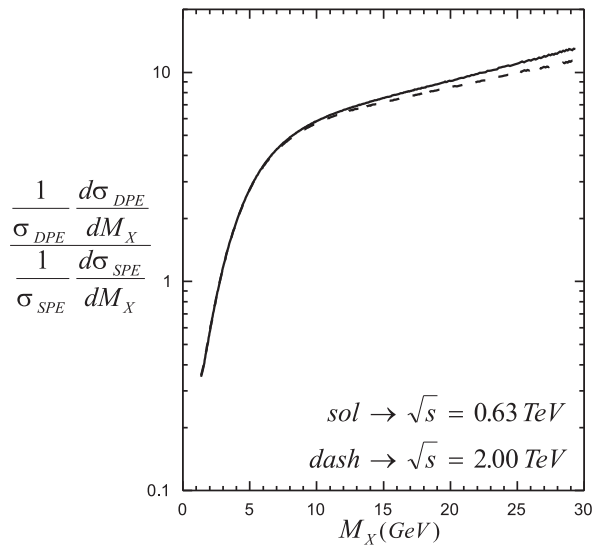,width=8cm}}
\caption{Ratio double/single Pomeron exchange mass distributions
as a function of $M_X$. In both cases we have assumed
$\sigma_{I\!\!P I\!\!P}= 1.0\,mb$ (for DPE processes) and
$\sigma_{p I\!\!P}=1.0\,mb$ (for SPE processes).}
\label{dndmxdperatio}
\end{center}
\end{figure}

\vspace{-0.7cm}

Although the normalization of our curves is arbitrary, the comparison
of the shapes  reveals a  difference between the two predictions.
Whereas the points  (from \cite{Brandt02}) show spectra broadening
with the c.m.s. energy, we predict  (solid line) the opposite
behavior: as the energy increases we observe a (modest) narrowing for
$d\sigma/dM_X$. {\it This small effect means that the diffractive
mass becomes a smaller  fraction of the available energy $\sqrt{s}$.}
In other words, the ``double diffractive inelasticity" decreases with 
energy in the same way as the ``diffractive inelasticity", as seen in 
Fig. \ref{xids}.

We are not able to make precise statements about the diffractive
cross section  (in particular about its normalization)  with our
simple model. Nevertheless,  the narrowing of $d\sigma_{DPE}/dM_X$
suggests a slower increase  (with $\sqrt{s}$)  of the integrated
distribution $\sigma_{DPE}$. We found this same effect  also for
$\sigma_{SPE}$. This trend is welcome and is one of the possible
mechanisms  responsible for the suppression  of diffractive cross
sections at higher energies  relative to some Regge theory predictions.

\begin{figure}[h]
\begin{center}
\centerline{\epsfig{figure=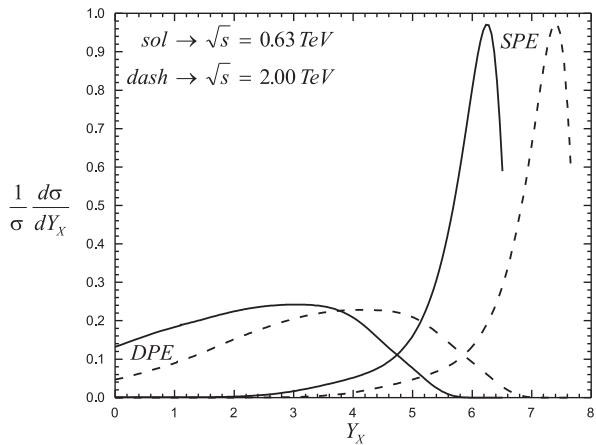,width=8cm}}
\caption{Double and single Pomeron exchange normalized rapidity
($Y_X$)  distributions. In both cases we have assumed
$\sigma_{I\!\!P I\!\!P}=1.0\,mb$  (for DPE processes) and
$\sigma_{p I\!\!P}=1.0\,mb$ (for SPE processes).} \label{dndydpe}
\end{center}
\end{figure}

In Fig. \ref{dndmxdperatio} we show the ratio $R(M_X)$ defined by:
\begin{equation}
R(M_X)= \frac{ \frac{1}{\sigma_{DPE}} \,
                      \frac{d \sigma_{DPE}}{d M_X} }
          {\frac{1}{\sigma_{SPE}} \, \frac{d \sigma_{SPE}}{d M_X}}
\label{ratioR}
\end{equation}
This quantity involves only distributions previously normalized to
unity and does  not directly compare the cross sections (which are
numerically very different for  DPE and single diffraction). In
$R$ the dominant $1/M^2_X$ factors cancel and we can
better analyse the details of the distributions which may contain
interesting  dynamical information. {\it The most prominent
feature of Fig. \ref{dndmxdperatio} is  the rise of the ratio with
$M_X$, almost by one order of magnitude in the mass range
considered. This can be qualitatively attributed to the fact that,
in single diffractive events the object $X$ has larger rapidities
than the corresponding  cluster formed in DPE events.} As a
consequence, when energy is released from the  incoming particles
in a SPE event, it goes more to kinetic energy of the $X$ system
(i.e., larger momentum $P_X$ and rapidity $Y_X$) and less to its
mass. In DPE,  although less energy is released, it goes
predominantly to the mass $M_X$ of the difractive cluster, which
is then at lower values of $Y_X$. In order to  illustrate this
behavior, we show in Fig. \ref{dndydpe} the rapidity distributions
of the $X$ (which has mass $M_X$) and $XY$ (which has mass
$M_{XY}$) systems. All curves are normalized to unity and with
them we just want to draw attention to the dramatically different
positions of the maxima of these distributions. The solid and
dashed lines show $ 1/\sigma \,  d \sigma / d Y_X $ for DPE
(curves on the left) and SPE (curves on the right) computed at
$\sqrt{s} = 630\,GeV$ and $\sqrt{s} = 2000\,GeV$, respectively. We
can clearly observe that DPE and SPE rapidity distributions are
separated by three units of rapidity and this difference stays
nearly constant as the c.m.s. energy increases. The location of
maxima in $ 1/\sigma \, d \sigma / d Y_X $ and their energy
dependence are predictions of our model.

\section{Leading particle spectra}

The leading particle effect is one of the most interesting features
of multiparticle production in hadron-hadron collisions.  In high
energy hadron-hadron collisions the momentum spectra of outgoing
particles which have the same quantum numbers as the incoming
particles, also called leading particle (LP) spectra, have been
measured already some time  ago \cite{Brenner82,Barton83}. Later on, new 
data on
pion-proton collisions were released  by the EHS/NA22 collaboration
\cite{EHS} in which the spectra of both  outcoming leading particles,
the pion and the proton, were simultaneously  measured. More recently
data on leading protons produced in eletron-proton  reactions at HERA
with a c.m.s. energy one order of magnitude higher than  in the other
above mentioned hadronic experiments became available  \cite{Cart}.
In the case of photoproduction,  data can be interpreted in terms of
the Vector Dominance Model \cite{VDM} and can therefore be considered
as data on LP production in vector meson-proton collisions. These new
measurements of LP spectra both in hadron-hadron  and in
eletron-proton collisions have renewed the interest on the subject,
specially because the latter are measured at higher  energies and
therefore the energy dependence of the LP spectra  can now  be
determined.

It is important to have a very good understanding of these spectra
for a  number of reasons. They are the input for calculations of the
LP spectra in hadron-nucleus collisions, which are a fundamental
tool in the description of atmospheric cascades initiated by cosmic
radiation   \cite{Cosmic,FGS}. There  are several new projects in cosmic ray
physics including  the High Resolution Fly's Eye Project, the
Telescope Array Project and the Pierre Auger Project \cite{CR} for
which a precise knowledge of energy flow (LP spectra  and
inelasticity distributions) in very high energy collisions  would be
very useful.

In a very different scenario, namely in high energy heavy ion
collisions at RHIC, it is very important to know where the outgoing
(leading) baryons are located in momentum space. If the stopping is
large  they will stay in the central rapidity region and affect the
dynamics there,  generating, for example, a baryon rich equation of
state. Alternatively,  if they populate the fragmentation region, the
central (and presumably hot and dense) region  will be dominated by
mesonic degrees of freedom. The composition of the dense matter is
therefore relevant for the study of quark gluon plasma  formation
\cite{QGP}.

In any case, before modelling $p - A$ or $A - A$ collisions one has
to  understand properly hadron-hadron processes. The LP spectra are
also  interesting for the study of diffractive reactions, which
dominate the  large $x_F$ region.

Since LP spectra are measured in reactions with low momentum transfer
and go up to large $x_F$  values, it is clear that the processes in
question occur in the  non-perturbative domain of QCD.  One needs
then ``QCD inspired''  models and the most popular are string models,
like FRITIOF, VENUS or the Quark Gluon String Model (QGSM).
Calculation of LP spectra involving these  models can be found in
Refs. \cite{BER} and \cite{WAL}.

\subsection{Leading particles in hadron-hadron collisions}

In the framework of the QCD parton model of high energy collisions,
leading particles originate from the emerging fast partons of the
collision debris. There is a large rapidity separation between fast
partons and  sea partons. Fast partons interact rarely with the
surrounding wee  partons. The interaction between the hadron
projectile and the target is  primarily through wee parton clouds. A
fast parton  or a coherent  configuration of fast partons may
therefore filter through essentially unaltered. Based on these
observations and aiming to study  $p - A$ collisons, the  authors of
Ref.  \cite{BER} proposed a mechanism for LP production in which  the
LP spectrum is given by the convolution of the parton momentum
distribution in the projectile hadron with its corresponding
fragmentation  function into a final leading hadron. This independent
fragmentation scheme is, however, not supported by leading charm
production in pion-nucleus  scattering. It fails specially in
describing the $D^{-}/D^{+}$ asymmetry. A number of models addressed
these data and the conclusion was  that valence quark recombination
is needed.  {\it Translated to leading pion or proton production this
means that what happens is rather a coalescence of valence quarks to
form the LP and not an independent fragmentation of a quark or
diquark to a  pion or a nucleon.} Another point is that the coherent
configuration formed by  the valence quarks may go through the target
but, due to  the strong stopping  of the gluon clouds, may be
significantly decelerated. This correlation  between central energy
deposition due to gluons   and  leading particle spectra was shown to
be essential for the undertanding of leading charm production
\cite{Duraes95}.

We follow the same general ideas of Ref. \cite{BER} but with a
different implementation. In particular we replace independent
fragmentation  by valence quark recombination and free leading parton
flow by deceleration due to ``gluon stripping''.

We have studied all measured LP spectra including those measured at
HERA. We have found some universal aspects in the  energy flow
pattern of all these reactions. Universality means, in the  context
of the IGM, that the  underlying dynamics is the same both in
diffractive and non-diffractive LP production and both in
hadron-hadron and photon-hadron processes.

In Ref. \cite{Duraes98a} we analyzed leading particle spectra in
hadronic collisions   and, assuming VDM, the leading proton
spectra in $e-p$ reactions. We have also  considered the
contribution coming from the diffractive processes. The leading
particle can emerge from different regions of the phase space, 
according to the values assumed by $x_{max}$ and $y_{max}$ in
eqs. (\ref{eq:OMEGAS}) and (\ref{eq:OMEGAH}).  The distribution
of its momentum fraction $x_L$ is given by:
\begin{eqnarray}
F(x_L)=(1-\alpha)\,F_{nd}(x_L) + \sum_{j=1,2}\alpha_j \,
F_{d}(x_L) \label{eq:FXLP}
\end{eqnarray}
where $\alpha = \alpha_1 + \alpha_2$ is the total fraction of
single diffractive  $(d)$ events from the lower and upper legs in
Fig. \ref{igmscenarios}, respectively.

\begin{figure}[h]
\begin{center}
\centerline{\epsfig{figure=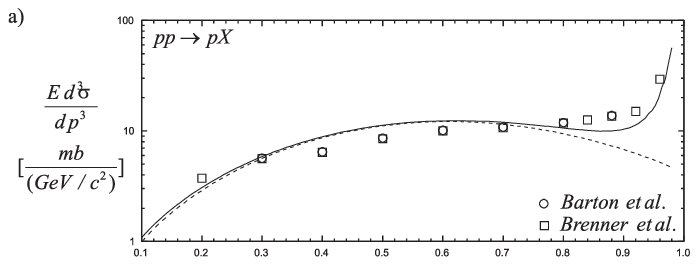,width=8.5cm}}
\label{igmlp1}
\end{center}
\end{figure}

\vspace{-1.1cm}

\begin{figure}[h]
\begin{center}
\centerline{\epsfig{figure=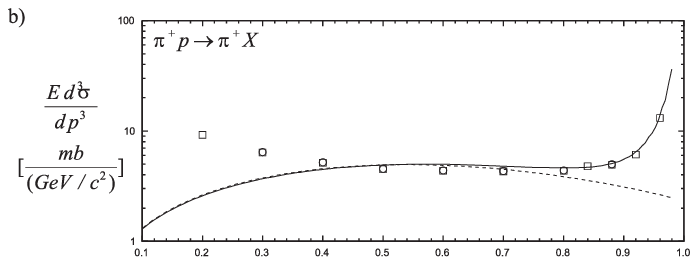,width=8.5cm}}
\label{igmlp1}
\end{center}
\end{figure}

\vspace{-1.1cm}

\begin{figure}[h]
\begin{center}
\centerline{\epsfig{figure=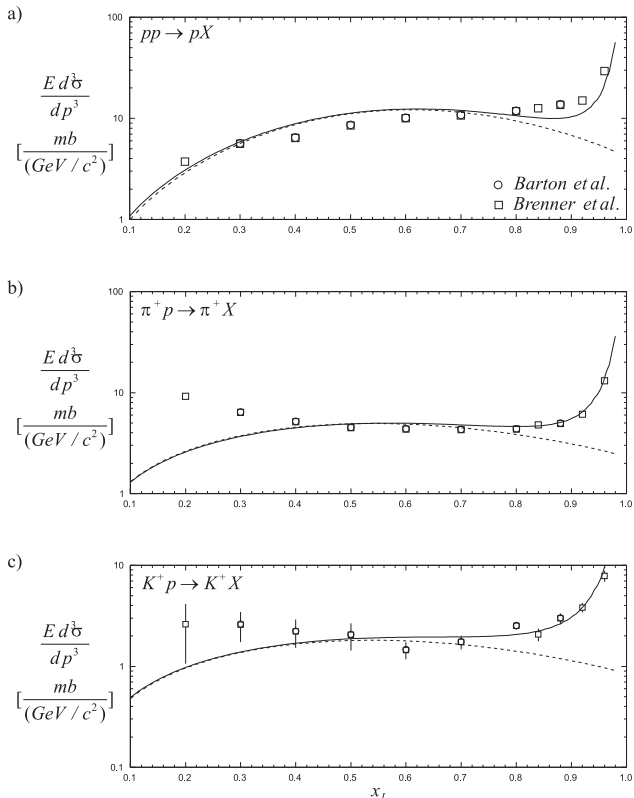,width=8.5cm}}
\caption{Comparison of our LP spectra $F(x_L)$ with data from
\protect\cite{Barton83} and \protect\cite{Brenner82}.}
\label{igmlp1}
\end{center}
\end{figure}


Notice that  $\alpha$ is essentially a new parameter here, which
should  be of the order of the ratio between the total diffractive
and total inelastic  cross sections \cite{Duraes97a}.

In Fig. \ref{igmlp1} we present our spectra of leading protons, pions
and kaons respectively. The dashed lines show the contribution of
non-diffractive LP production  and the solid lines show the effect of
adding a diffractive component,  calculated with the intermediate
Pomeron profile. All  parameters were fixed previously and the only
one to be fixed was $\alpha$. For simplicity we have neglected the
third diagram in Fig. 1 (c), because it gives a curve which is very
similar in shape to the non-diffractive curve. In contrast, the
Pomeron emission by the projectile (Fig. 1b) produces the diffractive
peak. We have then chosen $\alpha_2=0$ and $\alpha_1=\alpha=0.3$ in
all collision types.

As expected, the inclusion of the diffractive component flattens
considerably the  final LP distribution bringing it to a good
agreement with the available experimental  data \cite{Brenner82,Barton83}. 
In our
model there is some room for changes leading to fits with better
quality. We could, for example, use a prescription for hadronization
(as we did before  in \cite{Duraes94})) giving a more important role
to it, as done in Ref. \cite{BER}.  In doing this, however, we loose
simplicity and the transparency of the physical  picture, which are
the advantages of the IGM. We prefer to keep simplicity and
concentrate on the interpretation of our results. In first place it
is interesting  to observe the good  agreement between our curve and
data for protons  (Fig. \ref{igmlp1}a) in the low $x_L$ region. The
observed protons could have been  also centrally produced, i.e., they
could come from the CF. However we fit data  without the CF
contribution. This suggests, as expected, that all the protons in
this $x_L$ range are leading, i.e., they come from valence quark
recombination. In  Figs. \ref{igmlp1}b) and \ref{igmlp1}c) we observe
an excess at low $x_L$. This is  so because pions and kaons  are
light and they can more easily be created from the sea  (centrally
produced). Our distributions come only from the leading  jet  and
consequently pass below the data points. {\it A closer look into the
three dashed lines  in Fig. \ref{igmlp1} shows that pion and kaon
spectra are softer than the proton one.  The former peak at $x\simeq
0.56$ while the latter peaks at   $x\simeq 0.62$.}  In the IGM this
can be understood as follows. The energy fraction that goes to the
central fireball, $K=\sqrt{x y}$, is controled by the behaviour of
the function $\chi(x,y)^{nd}$, which is approximately a double
gaussian in the variables $x$ and  $y$, as it can be seen in
expression (\ref{eq:CHI}). The quantities $\langle x  \rangle$ and
$\langle y \rangle$ play the role of central values of this gaussian.
Consequently when $\langle x \rangle$ or $\langle y \rangle$
increases, this means  that the energy deposition from the upper or
lower leg (in Fig. 1) increases  respectively. The quantities
$\langle x \rangle$ and $\langle y \rangle$ are the moments of the
$\omega$ function and are directly  proportional to the gluon
distribution functions in the projectile and target and inversely
proportional to the target-projectile inelastic cross section. In the
calculations, there  are two changes when we go from $p-p$ to $\pi-p$:
\begin{itemize}
\item[$(i)$]
The first  is that we replace $\sigma^{pp}_{inel}$ by $\sigma^{\pi
p}_{inel}$ which is smaller. This leads to an overall increase of the
energy deposition. There are some indications that this is really the
case and the inelasticity in $\pi-p$  is larger than in $p-p$
collisions \footnote{For example, in the cosmic ray experiments
it is usually assumed that $K_{\pi N} = 1.5\, K_{pp}$, which is
traced to analysis of data like those in \cite{CRINEL} performed in
terms of the additive quark models (cf., \cite{CRINEL1}).}.

\item[$(ii)$] The second and most important change is that we replace
one gluon distribution in the proton $G^p(y)$ by the corresponding
distribution in the pion $G^{\pi}(y)$. We know that  $G^p(y) \simeq
(1-y)^5 /y$ whereas $G^{\pi}(y) \simeq (1-y)^2/y$, i.e., that gluons
in pions are harder than  in protons. This introduces an asymmetry in
the moments $\langle x \rangle$ and  $\langle y \rangle$, making the
latter significantly larger.
\end{itemize}

{\it As a consequence, because of their harder gluon distributions,
pions will be more  stopped and will emerge from the collision  with
a softer $x_L$ spectrum.} This can already be seen in the data points
of Fig.  \ref{igmlp1}. However since these points contain particles
produced by other  mechanisms, such as central and diffractive
production, it is not yet possible  to draw firm conclusions.  
One should mention here that there is another
possible difference between nucleons and mesons which can contribute
to the different behaviour of the leading particles in both cases. It
is connected with the triple gluon junction present 
in baryons but not in mesons, which, if treated as an elementary object,
can influence sunbstantially LP spectra (cf. \cite{FOOT}). We shall
not discuss this possibility in this paper.

The analysis of the moments $\langle x \rangle$ and $\langle y
\rangle$ can also be done for the diffractive process shown in
Fig. 1b).  Because of the  cuts in the integrations in eq.
(\ref{eq:defMOM}), they will depend on $x_L =1-y$. We calculate
them for $p+p \rightarrow p + X $ and $\pi + p \rightarrow \pi +
X$ reactions. For low $x_L$ they assume very similar values as in
the non-diffractive  case. For large $x_L$ however we find that
$\langle x \rangle_{p} \simeq \langle x \rangle_{\pi}$ and
$\langle y \rangle_{p} \simeq \langle y \rangle_{\pi}$. The reason
for these approximate equalities is that in diffractive processes
we cut  the large $y'$ region and this is precisely where the pion
and the proton would  differ, since only for large $y$ are
$G_{I\!\!P}^p(y) \simeq (1-y)^5 /y$ and $G_{I\!\!P}^{\pi}(y)
\simeq (1-y)^2/y$ significantly different. In Ref.
\cite{Duraes97a}  we have shown that the introduction of the above
metioned cuts drastically  reduces the energy ($\sqrt{s}$)
dependence of the diffractive mass distributions leading, in
particular, to the approximate $1/M^2_X$ behaviour for all values
of $\sqrt{s}$ from ISR to Tevatron energies. {\it Here these cuts
produce another type of scaling, which may be called ``projectile
scaling'' or ``projectile  universality of the diffractive peak''
and which means that for large enough  $x_L$ the diffractive peak
is the same for all projectiles.} The corresponding $\chi^{d}$
functions will be the same for protons and pions in  this region.
The cross section appearing in the denominator of the moments
will, in this case, be the same, i.e., $\sigma^{I\!\!P p}$.

The only remaining difference between them, their different
gluonic distributions, is in this region cut off. This may be
regarded as a prediction of the IGM. Experimentally this may be
difficult to check since one would need a large number of points
in large $x_L$ region of the leading particle spectrum. Data
plotted in Fig. \ref{igmlp1} neither prove nor disprove this
conjecture. The discrepancy observed in the proton spectrum is
only due to our choice of normalization of the diffractive and
non-diffractive curves. The peak shapes are similar.

\begin{figure}[h]
\begin{center}
\centerline{\epsfig{figure=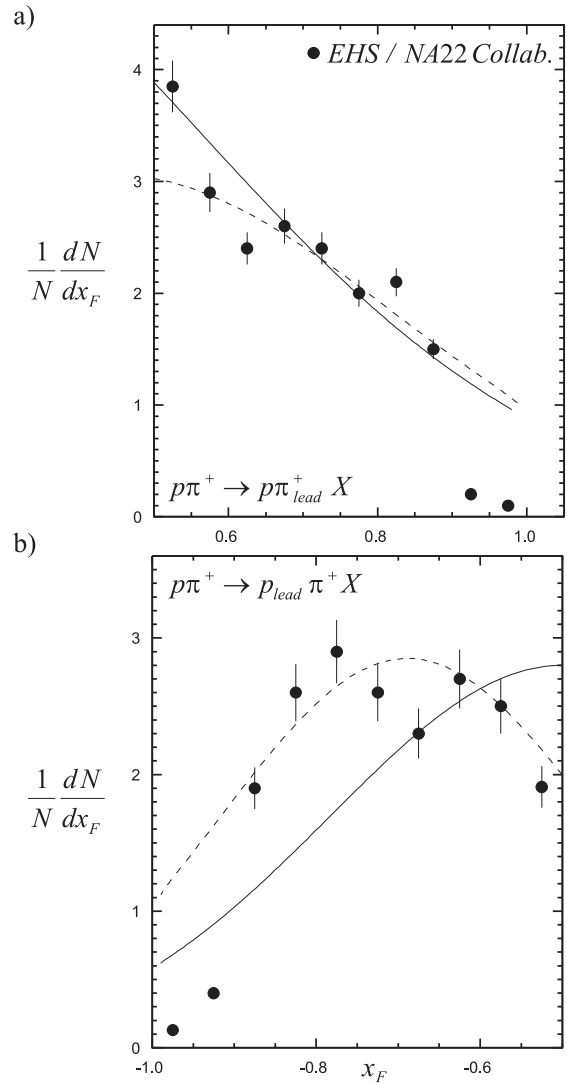,width=7.5cm}}
\caption{a) Comparison of our spectra $F(x_L)$ for leading pions
with data from Ref. \protect\cite{EHS} in the reaction
$\pi^{+} + p \rightarrow \pi^{+} + p + X$. Solid and dashed
lines correspond to the choices $m_0=0.35$ GeV and
$m_0=0.45$ GeV respectively. b) the same as a) for the
leading proton spectrum $F(x_F)$ measured in the same
reaction.}
\label{igmlp11}
\end{center}
\end{figure}
\vspace{-0.7cm}

The EHS/NA22 collaboration provided us with data on $\pi^{+}+p
\rightarrow \pi^{+}+p+X$ reactions. In particular they present the
$x_F$ distributions of both leading particles, the pion and the
proton. Their points for pions and protons are shown in Fig.
\ref{igmlp11}a) and b) respectively. These points are presumably
free from diffractive dissociation.  The above mentioned asymmetry
in pion and proton energy loss emerges clearly,  the pions being
much slower. The proton distribution peaks at $x_F\simeq 0.6-0.8$.
Our curves (solid lines) reproduce with no free parameter this
behaviour and we  obtain a good agreement with the pion spectrum.
Proton data show an excess at large $x_F$ that we are not able to
reproduce keeping the same values of parameters as  before.

The authors of Ref. \cite{EHS} tried to fit their measured proton
spectrum with the FRITIOF code and could not obtain a good
description of data. This indicates that  these large $x_F$ points
are a problem for standard multiparticle production models as
well. In our case, if we change our parameter $m_0$ from the usual
value $m_0=0.35\,GeV$ (solid line) to $ m_0=0.45\,GeV$ (dashed
line) we can reproduce most  of data points both for pions and
protons as well. This is not a big change and  indicates that the
model would be able to accomodate this new experimental
information. Of course, a definite statement about the subject
would require a global refitting procedure, which is not our main
concern now.

\subsection{Leading particles in photon-proton collisions}

If, at high energies, the reactions $\rho - p$ and $\pi - p$ have the
same characteristics and if VDM is a good hypothesis, then more about
the energy flow in meson-$p$ collisions can be learned at HERA.
Indeed, as mentioned in \cite{HERA}, at the HERA electron-proton
collider the bulk of the cross section corresponds to
photoproduction, in which a beam electron is scattered through a very
small angle and a quasi-real photon interacts with the proton. Using
VDM, high energy photoproduction  exhibits  therefore similar
characteristics to  hadron-hadron interactions.

Data taken by the ZEUS collaboration at HERA \cite{Cart} show that
the LP spectra measured in photoproducion and in DIS (where $Q^2 \geq
4  \,\,GeV^2$) are very similar, specially in the large $x_L$ region.
This suggests that, as pointed out in \cite{SNS}, the QCD hardness
scale for particle production in DIS gradually decreases from a
(large) $Q^2$,  which is relevant in the photon fragmentation region,
to a soft scale in the proton fragmentation region, which is the one
considered here. We can therefore expect a similarity of the
inclusive spectra of the leading protons in high energy hadron-proton
collisions, discussed above, and  in virtual photon-proton
collisions. In other words, we may say that the photon is neither
resolving nor being resolved by the fast  emerging protons. This
implies that these reactions are dominated by some  non-perturbative
mechanism. This is confirmed by the failure of perturbative QCD
\cite{AH}, (implemented by the Monte Carlo codes ARIADNE and HERWIG)
when applied  to the proton fragmentation region. In Ref. \cite{SNS}
the LP spectra were  studied in the context of meson and Pomeron
exchanges. Here we use the vector meson dominance hypothesis and
describe leading proton production in the same way as done for
hadron-hadron collisions. The only  change is that now we have
$\rho-p$ instead of $p-p$ collisions. Whereas this may be generally
true for photoproduction, it remains an approximation for DIS, valid
in the large $x_L$ region.

Assuming that VDM is correct, the incoming photon line can be
replaced by solid line in Fig. 1. During the interaction the photon
is converted into a  hadronic state, called $V$, and then interacts with the
incoming proton. At HERA only collisions  $V-p$ are
relevant. The state $V$ looses fraction $x$ of its original  momentum
and gets excited carrying a $x_F= 1 -x$ fraction of the initial
momentum. The proton, which we  call here the diffracted proton,
looses only a fraction $y$ of its momentum but otherwise remains
intact. We  assume here, for simplicity, that  the vector meson
is a $\rho^{0}$ and take  $ G^{\rho^{0}}(x) = G^{\pi}(x)$ in eqs.
(\ref{eq:OMEGAS}) and (\ref{eq:OMEGAH}).

\begin{figure}[h]
\begin{center}
\centerline{\epsfig{figure=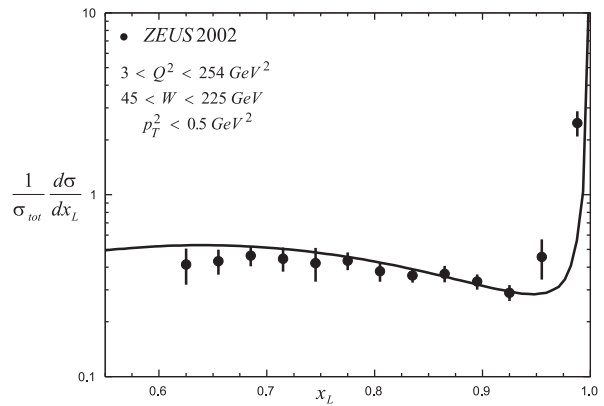,width=8cm}}
\caption{Comparison between our calculation and the  data on the
leading proton spectrum  measured at HERA by the ZEUS Collab.
\protect\cite{Garf03}.}
\label{igmlp2}
\end{center}
\end{figure}

\vspace{-0.7cm}

In Fig. \ref{igmlp2} we present our spectrum of leading protons in
$\gamma p$  collisions. All
parameters leading to the results in that figure are the same as
established before in our study  of diffractive mass distributions in
photon-proton collision at HERA.

\subsection{Leading $J/\psi$ production}

All produced particles come essentially from the gluons and
quark-antiquark pairs already pre-existing in the projectile and
target,  or radiated during the collision. This qualitative picture
takes different implementations in the many existing multiparticle
production models. In the IGM,  the produced particles (and
consequently the energy released in the secondaries and  lost by the
projectiles) come almost entirely from the pre-existing gluons in the
incoming hadrons. This conjecture may be directly  tested using a
high energy, nearly  gluonless hadronic projectile. In this case,
according to the IGM, inspite of the high  energy involved, the
production of secondaries would be suppressed  in comparison to  the
production observed in reactions induced by ordinary hadrons.  The 
energy would be  mostly carried away by the projectile leading
particle which would then be observed  with a hard $x_F$ spectrum.
This type of gluonless projectile is available in $J/\psi$
photoproduction, where the photon can be understood as a virtual
$c\bar c$ pair which  reacts with the proton and turns into the
finally observed $J/\psi$. There are low  energy data taken by the
FTPS Collaboration \cite{ftps} and  high energy data  from HERA
\cite{JPSIDATA}.

We want to stress here the fact that the fair agreement with data
observed in  Figs. (\ref{igmlp1}a) and (\ref{igmlp2}) is 
possible only because the diffraction processes  have been
properly incorporated in the calculations. In other words, the
inclusion  of a diffractive component turns out to be a decisive
factor to get agreement with data.  We can also describe
reasonably well pionic and kaonic LP and the observed difference
turns out to be due to their different gluonic distributions.

\begin{figure}[h]
\begin{center}
\centerline{\epsfig{figure=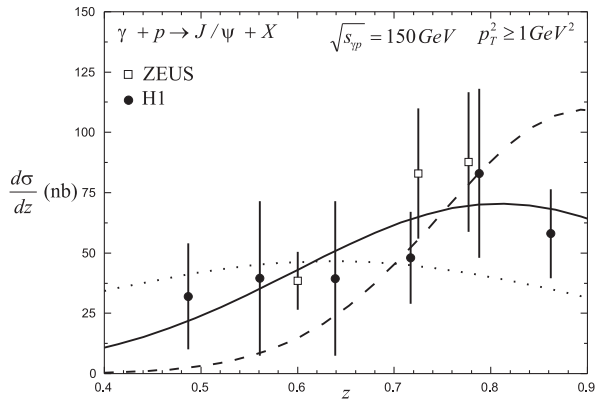,width=8cm}}
\caption{Comparison of the IGM distribution $F(z)$ with data of
Ref. \protect\cite{JPSIDATA} with restricted acceptance $p_T^2
\geq 1\, (GeV/c)^2$ and $0.5 \leq z \leq 0.9$ for fixed value of
$\sigma^{inel}_{J/\psi-p}=9\, mb$ and for three different values
of $p^{J/\psi}$: $0.066$ (dashed line), $0.033$ (solid line) and
$0.016$ (dotted line).} \label{igmlpjpsi}
\end{center}
\end{figure}

\vspace{-0.7cm}

The crucial role played by the parameter $p$ (see eq.
(\ref{eq:gdx})) representing the energy-momentum fraction of a
given hadron allocated to gluons is best seen in Fig.
\ref{igmlpjpsi} where we show the fit to data for leading $J/\psi$
photoproduction \cite{JPSIDATA}. The only parameter to which results 
are really sensitive 
is $p=p^{J/\psi}$ which, as shown in Fig. \ref{igmlpjpsi}, has to be
very small, $p^{J/\psi} = 0.033$. This is what could be expected from
the fact that charmonium is a non-relativistic system and almost
all  its mass comes from the quark masses leaving therefore only a 
small fraction,
\begin{equation}
p^{J/\psi}\, =\, \frac{M_{J/\psi} - 2 m_c}{M_{J/\psi}}\, \simeq \,
0.033 ,\label{eq:JPSIi}
\end{equation}
for gluons (here $m_c = 1.5\, GeV$ and $M_{J/\psi} = 3.1\, GeV$). Of
course, the value of $p^{J/\psi}$ required to give a very good fit of
data might change either  with another choice of $m_c$ or another
choice of  $\sigma^{inel}_{J/\psi-p}$. However  these changes might
affect $p^{J/\psi}$ by, at most,  a factor two. {\it  This suggests
that the momentum fraction carried by gluons in the $J/\psi$ is one
order of magnitude  smaller than that carried by gluons in light
hadrons.}

\section{Summary and conclusions}

We were able to fit an impressive amount of experimental 
data, which had nothing in common except the fact that they always 
referred to the momentum (or rapidity) distribution of some observed 
particle or to the invariant mass distribution of a cluster of 
measured particles. We could fit these data starting from  one single 
"generating" function, $\chi(x,y)$, which depends almost only on the 
density and interaction cross section of the gluons inside hadrons.
These are fundamental quantities in QCD and with our model we can test
the existing results for them. More than just fitting, we did some 
predictions and one of them, the leading particle spectrum shown in 
Fig. \ref{igmlp2}, was confirmed by experiment.

After all these works, we may ask ourselves what have we learned. We 
believe that we have constructed a simple and consistent picture of 
energy flow in strong interactions, based on the {assumption that 
{\it energy loss and leading particle spectra are determined by  many 
independent gluon-gluon collisions and valence quarks play a secondary 
role}. Consequently, energy flow will reflect the properties of the gluon
distributions and cross sections in the colliding hadrons. 
This picture seems to be universal, i.e., valid in many different 
contexts. However, in order to see this universality we have  to be careful and  
use proper kinematical limits of the phase space for every reaction 
considered, as illustrated in Figs. \ref{igmscenarios} and 
\ref{igmphasespace}. When
this is done the sensitivity  of energy flow to  other (than gluon 
distributions and cross sections) aspects of the
production process is only of secondary importance and needs special
observables (which are sensitive to, for example, the quantum numbers of
the detected particles) to be visible. But even then,  the IGM is 
indispensable because it provides the important energy correlations between
different parts of the phase space. 

Our analysis shows also clearly that our model can be regarded as a 
useful reference point for all more sophisticated approaches whereas, 
for hydrodynamical approaches of multiparticle production,  it provides the 
 initial energy used for the further evolution and hadronization of the
created systems. However, in order to comply
with the  recent developments of QCD concerning the low $x$ gluonic 
content of hadrons \cite{raju,dumitru} it must  be accordingly updated. 
We plan to do this in the future. We also plan to account for the
intrinsic fluctuations present in the hadronizing systems. In the
usual statistical models this can be done by using the so called
nonextensive statistics and, as was shown in \cite{qK}, it can
influence substantially some energy flow results, in particular the
estimation of inelasticity $K$.

\appendix
\section{}

\vspace{0.5cm}
\noindent
{\bf The main ideas}
\vspace{0.5cm}

The IGM can be summarized in the following way:
\begin{itemize}
\item[$(i)$] The two colliding hadrons are represented by valence
quarks carrying their quantum numbers  plus the accompanying
clouds of gluons.

\item[$(ii)$] In the course of a collision the gluonic clouds
interact strongly depositing in the central region of the reaction
fractions $x$ and $y$ of the initial energy-momenta of the respective
projectiles in the form of a gluonic {\it Central Fireball} (CF).

\item[$(iii)$] The valence quarks get excited and form 
{\it Leading Jets} (LJ's)
which decay and  populate mainly the fragmentation regions of the
reaction.
\end{itemize}

The fraction of energy stored in the CF is therefore equal to
$K\, =\, \sqrt{x y}$ and its rapidity is $Y=\frac{1}{2}\,
\ln \frac{x}{y}$.

The CF consists of {\it minifireballs} (MF)
formed from pairs of colliding gluons. In collisions at higher scales
a MF is the same as a pair of minijets or jets. In the study of
energy flow the details of fragmentation and hadron production are
not important. Most of the MF's will be in the central region and we
assume that they coalesce forming the CF. The collisions leading to
MF's occur at different energy scales given by $Q^2_i = x_i\,y_i\,s$,
where the index $i$ labels a particular kinematic  configuration
where the gluon from the projectile has momentum $x_i$ and  the gluon
from the target has $y_i$. We have to choose the scale where we start
to use perturbative QCD.  Below this
value we have to assume that we can still talk about individual soft
gluons and due to the short correlation length  
between them  they still interact mostly pairwise. In
this region we can no longer use  the distribution functions
extracted from DIS nor the perturbative elementary cross sections.

\vspace{0.5cm}
\noindent
{\bf The central formula}
\vspace{0.5cm}

The central quantity in the IGM is the probability to form a CF
carrying momentum fractions $x$ and $y$ of two colliding hadrons. It
is defined as the sum over an undefined number n of MF's:
\begin{eqnarray}
\chi(x,y) & = & \sum_{n_1} \, \sum_{n_2} \cdots \sum_{n_i} \, \delta
  \left[ x - n_1 \, x_1 - \cdots - n_i \, x_i \right] \nonumber \\
&\times& \, \delta \left[
   y - n_1 \, y_1 - \cdots - n_i \, y_i \right] \, P(n_1) \cdots
    P(n_i) \nonumber \\
& = & \sum_{\{n_i\}} \left\{ \delta \left[ x - \sum_i n_i\, x_i \right]
   \delta \left[ y - \sum_i n_i\, y_i \right] \right\} \nonumber \\
&\times& \prod_{\{n_i\}} P(n_i)
\label{eq:isachi}
\end{eqnarray}

The delta functions in the above formula garantee energy momentum
conservation and $P(n_i)$ is the probability to have $n_i$ collisions
between gluons with $x_i$ and $y_i$. The expression above is quite general.
It becomes specific when we define $P(n_i)$.  
The assumption of multiple parton-parton incoherent scattering 
(which is also used in Refs. \cite{wang,brown,gs,sj,sapeta}) implies 
a  Poissonian distribution of the number of  parton-parton collisions and   
thus $P(n_i)$ is given by:
\begin{eqnarray}
P(n_i) &=& \frac{(\overline n_i)^{n_i} \, exp(-\overline n_i)}{n_i\,!}
\end{eqnarray}

Inserting $P(n_i)$ in (\ref{eq:isachi}) and using the following
integral  representations for the delta functions:
\begin{eqnarray}
&&\delta \left[ x - \sum_i n_i \, x_i \right] \, = \nonumber \\
&=& \frac{1}{2\pi}
     \int^{+\infty}_{-\infty} dt \,\exp \left[ it \left( x - \sum_i
      n_i \, x_i \right)\right]
\end{eqnarray}
\begin{eqnarray}
&&\delta \left[ y - \sum_i n_i \, y_i \right] \, = \nonumber \\
&=&  \frac{1}{2\pi}
    \int^{+\infty}_{-\infty} du\, \exp \left[ iu \left( y - \sum_i
     n_i \, y_i \right)\right]
\end{eqnarray}
we can perform all summations and products arriving at:
\begin{eqnarray}
\chi(x,y) \, &=& \, \frac{1}{(2\pi)^2} \int^{+\infty}_{-\infty} dt
    \int^{+\infty}_{-\infty} du \,\exp \,[i (tx + uy)]  \nonumber \\
&\times&
\exp \left\{
     \sum_i \left\{ \overline{n}_i \left[ e^{-i(tx_i+uy_i)} - 1
      \right] \right\} \right\}
\label{eq:yf}
\end{eqnarray}

Taking now the continuum limit:
\begin{eqnarray}
\overline{n}_i \, = \, \frac{d \overline{n}_i}{dx' \, dy'} \,
    \Delta x' \Delta y'  \;\; \longrightarrow \; \; d\overline{n}
     \, = \, \frac{d\overline{n}}{dx' \, dy'} \, dx' \, dy'
\end{eqnarray}
we obtain:
\begin{eqnarray}
&&\chi(x,y)  =  \frac{1}{(2\pi)^2} \int^{+\infty}_{-\infty} dt
    \int^{+\infty}_{-\infty} du \,\exp[i(tx + uy)] \nonumber \\
   &\times& \exp\left\{ \int^1_0 dx' \int^1_0 dy' \, \omega(x',y')
       \left[ e^{-i (tx' + uy')} -1 \right]\right\}
\label{eq:yfs}
\end{eqnarray}
where
\begin{eqnarray}
\omega(x',y')\,=\, \frac{d \overline {n}}{ d x' \, d y'} .
\end{eqnarray}

This function $\omega(x',y')$ is called the spectral
function and  represents the average number of gluon-gluon collisions as
a function of  $x'$ e $y'$. It contains all the dynamical inputs of the
model and has the form:
\begin{eqnarray}
\omega(x',y')\, &=&\, \frac{\sigma_{gg}(x'y's)}{\sigma(s)}
   \, G(x')\, G(y') \nonumber \\
&\times& \theta\left(x'y' - K^2_{min}\right),
   \label{eq:omega}
\end{eqnarray}
where $G$'s denote the  gluon distribution functions in  the
corresponding projectiles and $\sigma_{gg}$ and $\sigma$ are the
gluon-gluon and hadron-hadron cross sections, respectively. In the
above expression $x'$ and $y'$ are the fractional momenta of two
gluons coming from the projectile and from the target whereas
$K_{min}=m_0 / \sqrt{s}$, with $m_0$  being the mass of lightest
produced state and $\sqrt{s}$ the total c.m.s. energy. $m_0$ is a
parameter of the model.

The integral in the second line of eq. (\ref{eq:yfs}) is dominated by
the  low $x'$ and $y'$ region. Considering the singular behavior of
the  $G(x)$  distributions at the origin we make the following
approximation:
\begin{equation}
e^{-i (tx' + uy')} -1\,\simeq\, -i  (tx' + uy') -\frac{1}{2}
(tx' + uy')^2
\end{equation}

With this approximation it is possible to perform the integrations in
(\ref{eq:yfs}) and obtain the final expression for $\chi(x,y)$
discussed  in the main text:
\begin{eqnarray}
&&\chi (x,y) =\frac{\chi _0}{2\pi \sqrt{D_{xy}}}\,
\exp\{-\frac 1{2D_{xy}}
[ \langle y^2\rangle (x-\langle x\rangle )^2 \nonumber \\
&+&\, \langle x^2\rangle (y-\langle y\rangle )^2
+\,2\,\langle xy\rangle
(x-\langle x\rangle )\,(y-\langle y\rangle )] \}
\label{eq:chiapp}
\end{eqnarray}
where
\begin{eqnarray}
D_{xy} &=& \langle x^2\rangle \langle y^2\rangle -
           \langle xy\rangle ^2  \nonumber
\end{eqnarray}
and
\begin{eqnarray}
\langle x^ny^m\rangle &=& \int_0^1\! dx\,x^n\, \int_0^1\! dy\,
y^m\, \omega (x,y), \label{eq:defmoms}
\end{eqnarray}
$\chi_0$ is a normalization factor defined by the condition:
\begin{eqnarray}
 \int_0^1\!dx\, \int_0^1\! dy\, \chi(x,y) \theta(xy - K_{min}^2) = 1
\end{eqnarray}

\vspace{0.5cm}
\noindent
{\bf The numerical inputs}
\vspace{0.5cm}

In order to evaluate the distribution  (\ref{eq:chiapp}) we need to
choose  the value of  $m_0$,  the semihard scale $p_{T\,min}$ and
define   $G(x)$ and $\sigma_{gg}$ in both interaction regimes. We
take  $p_{T\,min}\,=\,2.3 \,GeV$ and $m_0\,=\, 0.35 \,GeV$. These are
the two scales present in the model. The semihard gluon-gluon cross
section is taken, at order $\alpha_s^2$, to be:
\begin{eqnarray}
\hat{\sigma}^h_{gg}(x\,,\,y\,,\,s)\,=\,
\kappa \frac{\pi}{16\,p^2_{T\,min}} \left[
\alpha_s(Q^2)\right]^2\,H
\end{eqnarray}
where
\begin{eqnarray}
H=36T\,+\frac{51\lambda T}{4\,x\,y}-
\frac{3\lambda^2\,T}{8\,x^2\,y^2}+\frac{9\lambda}{x\,y}\,
\ln\left[\,\frac{1-T}{1+T}\,\right]
\end{eqnarray}
and
\begin{eqnarray}
T\,=\,\left[1-\frac{\lambda}{x\,y}\right]^{\frac{1}{2}} \,;\,
\lambda\,=\, \frac{4\,p^2_{T\,min}}{s}
\end{eqnarray}
The parameter $\kappa$ is the one frequently used to incorporate
higher corrections in $\alpha_s$ and is $1.1 \leq \kappa \leq 2.5$
according to the choice of $G(x)$, of the scale $Q^2$ and
$p_{T\,min}$. For $p_{T\,min}=2.3 \,GeV$, $\kappa=2.5$.

The coupling constant is given by:
\begin{eqnarray}
\alpha_s(Q^2)\,=\, \frac{12\,\pi}{\left(33-2 N_f\right)\,
\ln\left[\frac{Q^2}{\Lambda^2}\right]}
\end{eqnarray}
where $\Lambda\,=\,0.2 \,GeV$ and $N_f\,=\,3$ is the number of active
flavors. As usual in minijet physics we choose $Q^2\,=\,p^2_{T\,min}$
and use the distributions $G(x,Q^2)$ parametrized in literature.

When the invariant energy of the gluon pair $\hat{s}$ is the interval
$m^2_0\,\leq\, \hat{s}\,=\,xys\,\leq\,4p^2_{T\,min}$ we are outside
the  perturbative domain. Parton-parton cross sections in the
non-perturbative regime  have been parametrized in \cite{valin}
leading to a successful quark-gluon model for  elastic and
diffractive scattering. Recently these non-perturbative cross
sections have been calculated in the stochastic vacuum model
\cite{menon}. The obtained cross sections are functions of the gluon
condensate and of the gluon field correlation  length, both
quantities extracted from lattice QCD calculations. In order to keep
our treatment simple we  adopt the older parametrization for the
gluon-gluon cross section used in \cite{valin}:
\begin{eqnarray}
\hat{\sigma}^s_{gg}(x\,,\,y\,,\,s)\,=\, \frac{\alpha} {x\,y\,s}
\end{eqnarray}
where $\alpha$ is a  parameter of the model \cite{IGM}.

\vspace{0.5cm}
\noindent
{\bf The main distributions}
\vspace{0.5cm}

Given $\chi(x,y)$ we can immediately write the inelasticity
distribution, its  complementary distribution, the leading jet
momentum spectrum and the CF rapidity  distribution:
\begin{eqnarray}
\chi(K) & = & \int^1_0 dx \int^1_0 dy \, \chi(x,y) \,
\delta\left(\sqrt{xy} - K \right) \nonumber\\
&\times&\,\theta\left(xy - K^2_{min} \right)
\label{eq:chidk}
\end{eqnarray}
\begin{eqnarray}
F(x_L)\, =\, &&(1-\alpha)\int^1_{x_{min}}\!\! dx\,
\chi^{(nd)}\left(x;y=1-x_L\right)\, + \nonumber\\
&& +\, \sum_{j=1,2}\alpha_j
\int^1_{x_{min}}\!\! dx\, \chi^{(d)}\left(x;y=1-x_L\right)
\label{eq:fxl}
\end{eqnarray}
\begin{eqnarray}
\chi(Y) & = & \int^1_0 dx \int^1_0 dy \, \chi(x,y)\,
\delta\left(\frac{1}{2}\, ln\,\left(\frac{x}{y} \right)- Y \right)
\nonumber\\
&\times&\,\theta\left(xy - K^2_{min} \right)
\label{eq:chidy}
\end{eqnarray}
where $\alpha = \alpha_1 + \alpha_2$ is the total fraction of single
diffractive $(d)$ events (from the upper and lower legs in Fig.
\ref{igmscenarios}, respectively) and where
\begin{equation}
x_{min} = {\rm Max}\left[\frac{m_0^2}{(1-x_L)s};
\frac{\left(M_{LP}+m_0)\right)^2}{s}\right]
\label{eq:xmina}
\end{equation}

The mass spectra for Single Diffractive processes are given by:
\begin{eqnarray}
\frac{1}{\sigma}\frac{d\sigma}{dM_X^2}&=&\frac{dN}{dM_X^2}=
\int_0^1dx\,\int_0^1dy\,\chi(x,y)\nonumber \\
&\times&\,\delta\left(M^2_X-sy\right)\,\theta\left(xy -
K_{min}^2\right) \label{eq:SDDa}
\end{eqnarray}

Substituting now eq. (\ref{eq:chiapp}) into eq. (\ref{eq:SDDa}) we
arrive at the following simple expression for the diffractive mass
distribution:
\begin{equation}
\frac{dN}{dM^2_X}\, =\, \frac{1}{s}\, F(M^2_X,s) \,H(M_X^2,s)
\label{eq:DIS}
\end{equation}
where
\begin{eqnarray}
F(M^2_X,s) &=&  \exp \left[ - \frac{\langle x^2\rangle}{2D_{xy}}\,
                 \left( \frac{M^2_X}{s} - \langle y\rangle \right) ^2
                 \right] \label{eq:FFF}
\end{eqnarray}
and
\begin{eqnarray}
H(M^2_X,s) = \frac{\chi_0}{2\pi \sqrt{D_{xy}}}\,
  \int^1_{\!_{\frac{m_0^2 }{M_X^2}}}\! dx\,\exp \left\{-\frac{1}{2D_{xy}}\,Z\right\}
    ,\label{eq:H}
\end{eqnarray}
$$
Z=[\langle y^2\rangle (x - \langle x\rangle )^2 -
               2\langle xy\rangle (x - \langle x\rangle)
               ( M_X^2/s - \langle y \rangle)]
$$

We first keep only the most singular parts of the gluonic
distributions used (i.e., $G(x)\simeq 1/x$) and collect all other
factors in eq. (\ref{eq:omega}) in a single parameter $c$. Assuming
that the ratio of the cross sections
$\frac{\sigma(xys)}{\sigma(s)}$ does not depend on $x$ and $y$ and
neglecting all terms of the order of $\frac{m_0^2}{s}$ and
$\frac{m_0^2}{M_X^2}$, we arrive at the following expressions for
the moments calculated in eq. (\ref{eq:defmoms}):
\begin{eqnarray}
\langle x\rangle \, &=& 2\, \langle x^2\rangle \simeq\,
                    c\, \ln \frac{M_X^2}{m_0^2}  \label{eq:momX}\\
\langle y\rangle \, &=& 2\, \frac{s}{M_X^2}\, \langle y^2\rangle
                       \simeq \, c\, \frac{M_X^2}{s}\,
                    \ln \frac{M_X^2}{m_0^2} \label{eq:newY}\\
\langle x\, y\rangle \, &\simeq& \, c\, \left(\frac{M_X^2}{s}\, -
           \frac{m_0^2}{s}\, \ln\frac{M_X^2}{m_0^2}\right)
            \label{eq:momXY}
\end{eqnarray}
Notice that in all cases of interest $\langle x\, y\rangle$ is
much smaller than other moments (by a factor
$\ln\frac{M_X^2}{m_0^2}$, at least). It means that $D_{xy}\,
\simeq\, \langle x^2\rangle\langle y^2\rangle $ and consequently
\begin{eqnarray}
F(M^2_X,s)\, &\simeq&\, \exp\left[ - \frac{\left(\frac{M_X^2}{s}\,
-\,
                       \langle y\rangle\right)^2}
                       {2\, \langle y^2\rangle}\right] \nonumber\\
         &\simeq&\, \exp\left[ - \frac{\left(1\, -\,
                        c\, \ln\frac{M_X^2}{m_0^2}\right)^2}
                        {c\, \ln\frac{M_X^2}{m_0^2}}\right]
\label{eq:F}
\end{eqnarray}
and
\begin{eqnarray}
H(M_X^2,s)\, &\simeq&\, \frac{\chi_0}{2\pi \sqrt{D_{xy}}}\,
                  \int^1_{\!_\frac{m_0^2}{M_X^2}}\! dx
                   \exp\left[ - \frac{(x - \langle x\rangle
                    )^2}{2\langle x^2\rangle}\right]\nonumber\\
            &\simeq&\, {\rm const}\,
            \frac{\sqrt{\langle x^2\rangle}}{\sqrt{D_{xy}}}
            \, =\, {\rm const}\,
            \frac{1}{\sqrt{\langle y^2\rangle}}\nonumber\\
             &\simeq&\, {\rm const}\, \frac{s}{M_X^2
                \, \sqrt{c\, \ln\frac{M_X^2}{m_0^2}}} \label{eq:newapprH}
\end{eqnarray}
leading to
\begin{eqnarray}
\frac{dN}{dM_X^2}\, &\simeq&\, \frac{1}{s}\, H(M_X^2,s)\,
                   F(M^2_X,s) \simeq\,\frac{\rm const}{M_X^2}\,\times
                   \nonumber\\
                    &\times&\frac{1}{\sqrt{c\, \ln\frac{M_X^2}{m_0^2}}}
                        \,
                        \exp\left[ - \frac{\left(1\, -\,
                        c\, \ln\frac{M_X^2}{m_0^2}\right)^2}
                        {c\, \ln\frac{M_X^2}{m_0^2}}\right]
                        \label{eq:result}
\end{eqnarray}

The mass spectra for Double Pomeron Exchange processes are given
by:
\begin{eqnarray}
\frac{1}{\sigma}\frac{d\sigma}{dM_{XY}}&=&\frac{dN}{dM_{XY}}=
\int_0^1 dx\,\int_0^1 dy\,\chi(x,y) \nonumber \\
&\times&\,\delta\left(M_{XY}-\sqrt{xys}\right)\,\theta\left(xy -
K_{min}^2\right) \label{eq:DPEa}
\end{eqnarray}

\section{}

\vspace{0.5cm}
\noindent
{\bf The  "kinematical"  Pomeron}
\vspace{0.5cm}

The Pomeron for us is just a collection of gluons which belong to the
diffracted proton or antiproton.  
These gluons behave like all other ordinary gluons  in
the proton and have therefore the same momentum distribution. The
only difference is the momentum sum rule,  which for the gluons in
$I\!\!P$  is:
\begin{eqnarray}
\int_0^1\! dx'\,x'\,G_{I\!\!P}(x')\,=\,p_d
\label{sumrule}
\end{eqnarray}
where $p_d=0.05$ (see \cite{Duraes97a})
instead of $p=0.5$, which holds for the entire gluon population in
the proton.

In order to make contact with the analysis performed by HERA
experimental groups we consider  two possible momentum distributions
for the gluons inside  $I\!\!P$. A hard one:
\begin{eqnarray}
f^h_{g/I\!\!P}(\beta) &=&  a_{h}\,(1\,-\,\beta)
\label{fhard}
\end{eqnarray}
and a ``super-hard''  (or ``leading gluon'') one:
\begin{eqnarray}
f^{sh}_{g/I\!\!P}(\beta) &=&  a_{sh}\, \beta^7 \, (1\,-\,\beta)^{0.3}
\label{superhard}
\end{eqnarray}
where $\beta$ is the momentum fraction of the Pomeron carried by the
gluons and the superscripts $h$ and $sh$ denote hard and superhard
respectively. The constants   $a_{h}$ and $a_{sh}$ will be fixed by
the sum rule (\ref{sumrule}).  In the past, following the same
formalism, we have also considered  a soft gluon distribution for the
Pomeron of the type
\begin{equation}
f^s_{g/I\!\!P}(\beta) =  6\,\frac{(1\,-\,\beta)^5}{\beta}
\label{supersoft}
\end{equation}
but  we found that this ``soft Pomeron'' distribution was
incompatible with  the single diffractive mass spectra measured at
HERA \cite{h1}. This Pomeron profile  was also ruled out by other
types of observables, as concluded in Refs. \cite{zeus} and
\cite{HERA}.

We use the Donnachie-Landshoff  Pomeron flux factor, which,
after the integration in the $t$ variable, is approximately given by
\cite{terron}:
\begin{equation}
f_{I\!\!P/p}(x_{I\!\!P}) \simeq  C \, x_{I\!\!P}^{1-2 \alpha_
{I\!\!P}}  \simeq   C \,\frac{1}{x_{I\!\!P}}
\end{equation}
where $x_{I\!\!P}$ is the fraction of the proton momentum carried
by the Pomeron and the normalization constant $C$ fixed also with
the help of (\ref{sumrule}). Noticing  that $\beta =
\frac{x}{x_{I\!\!P}}$, the distribution $G_{I\!\!P}(y)$ needed  in
eqs. (\ref{eq:OMEGAS}) and (\ref{eq:OMEGAH}) is then given by the
convolution:
\begin{eqnarray}
G^{s,h,sh}_{I\!\!P} (y) &=& \int_y^1\! \frac{ dx_{I\!\!P} }
{x_{I\!\!P}} \,
f_{I\!\!P/p}(x_{I\!\!P})\,f^{s,h,sh}_{g/I\!\!P}(\frac{y}{x_{I\!\!P}})
\label{conv}
\end{eqnarray}

We use also the ``diffractive gluon distribution'' given by:
\begin{equation}
G_{I\!\!P}(y) =  a\,\frac{(1\,-\,y)^5}{y}
\label{gFeynman}
\end{equation}
where $a$ is fixed by the sum rule.


\vspace{1.cm}
\underline{Acknowledgements}: This work has been supported by FAPESP,
CNPQ (Brazil) (FOD and FSN) and by KBN (Poland) (GW). We are indebted
to  our colleagues Y. Hama, T. Kodama, M. Menon, F. Grassi, S.
Padula, members  of the  ``Projeto Tem\'atico FAPESP'' and also to R.
Covolan and A. Natale, V.P Gon\c{c}alves for  fruitful discussions.

\end{document}